\documentclass[twocolumn,superscriptaddress,10pt,aps,prb]{revtex4-2}
\usepackage{epsfig,amsopn}
\usepackage{graphicx}
\usepackage{epstopdf}
\usepackage{braket}
\usepackage{amsmath,amssymb}
\usepackage{amsthm}
\usepackage{enumerate}
\usepackage{xcolor}
\newcommand*{\balancecolsandclearpage}{%
	\close@column@grid
	\cleardoublepage
	\twocolumngrid
}

\newcommand\la{\langle}
\newcommand\ra{\rangle}
\newcommand\al{\alpha}
\newcommand\bt{\beta}

\def\nodag{{\vphantom{dagger}}}
\usepackage[defaultcolor=blue]{changes}
\definechangesauthor[name=Babak, color=orange]{b}
\definechangesauthor[name=rekha, color=green]{gr}

\newcommand{\RNum}[1]{\uppercase\expandafter{\romannumeral #1\relax}}

\begin{document}
	\title{Josephson-Current Signatures of Unpaired Floquet Majorana Bound States}
	
	\author{Rekha Kumari}
	\affiliation{Department of Physics, Indian institute of technology, Kanpur, India}
	
	\author{Babak Seradjeh}
	\affiliation{Department of Physics, Indiana University, Bloomington, Indiana 47405, USA}
	\affiliation{Quantum Science and Engineering Center, Indiana University, Bloomington, Indiana 47405, USA}
	\affiliation{IU Center for Spacetime Symmetries, Indiana University, Bloomington, Indiana 47405, USA}
	
	\author{Arijit Kundu}
	\affiliation{Department of Physics, Indian institute of technology, Kanpur, India}
	
	\begin{abstract}
		We theoretically study the transport signatures of unpaired Floquet Majorana bound states in the Josephson current of weakly linked, periodically driven topological superconductors. We obtain the occupation of the Floquet Majorana modes in the presence of weak coupling to thermal leads analytically, and show that, similar to static superconductors, the Josephson current involving Floquet Majorana bound states is also $4\pi$-periodic in the phase difference across the junction, and also depends linearly on the coupling between superconductors. Moreover, unlike the static case, the amplitude of the Josephson current can be tuned by setting the unbiased chemical potential of the driven superconductors at multiple harmonics of the drive frequency. As a result, we uncover a \emph{Josephson Floquet sum rule} for driven superconductors. We confirm our analytical expressions for Josephson current, the occupation of Floquet bands, and a perturbative analysis of the quasienergies with numerically exact results.
	\end{abstract} 
	
	\maketitle
	
	\section{Introduction}%
	Majorana fermions~\cite{Ettore_1937,Kitaev_2001,Alicea_2012} are their own anti-particles that appear in condensed matter systems as quasiparticles with equal superposition of electrons and holes and are of immense importance for possible applications in fault-tolerant quantum information processing. Among many possible candidates for hosting such states, Majorana fermions as topologically protected edge states of one-dimensional topological superconductors have been studied extensively both theoretically~\cite{Wimmer_2010,Kane_2010,Zhang_2011,Linder_2010,Leijnse_2012,Potter_2010,Lutchyn_2011,Beenakker_2013,Stanescu_2013,MFranz_2015,Majref10,Majref11,Majref12,Majref13} and experimentally~\cite{zeropeak1exp,zeropeak2exp,zeropeak3exp,zeropeak4exp,zeropeak5exp,fracjos1exp,fracjos2exp,fracjos3exp} in recent years. These Majorana bound states have unique transport signatures: it is argued that quantized zero-bias conductance~\cite{zeropeak1,zeropeak2,zeropeak3,zeropeak4} as well as unusual 4$\pi$-periodic Josephson effect~\cite{fracjos1,fracjos2,fracjos3,Majref7,MajreF9} can identify their existence, leading to extensive search for Majorana modes in various solid-state systems.
	
	More recently, periodically driven quantum systems (often called Floquet systems) have been studied as a promising platform for realizing topologically non-trivial states by band-structure engineering~\cite{Floref1,Floref4,Floref5,Floref6,Floref7,Floref8,Floref9,Floref10,Floref11,Floref12,Floref13,Floref14,Floref15,Floref16,Floref17,Floref18} and there has been a surge in experimental activities in the search for topological states in solid state~\cite{flo_solid1,flo_solid2}, cold-atom~\cite{flo_cold1,flo_cold2,flo_cold3,flo_cold4,flo_cold5} and optical systems~\cite{flo_opt1,flo_opt2,flo_opt3}, which are driven periodically. It is argued that a periodically driven one-dimensional superconductors can host a number of \textit{far-from equilibrium} edge states, which have the same character of static Majorana fermions~\cite{zeropeak2a,floq_maj1,floq_maj2}. These `Floquet' Majorana fermions (FMFs) are the result of non-trivial topological nature of the underlying periodically driven superconductor, where the periodic drive can further tune the topological character of the state~\cite{Floquetref1,Floquetref2,Floquetref3,Floquetref4,Floquetref5,Floquetref6,Floquetref7,Floquetref9,Floquetref11,Floref2,Floref3}. Unlike their static counterparts, the FMFs appear in two flavors (with different quasienergies) and their occupations do not follow equilibrium distribution functions, leading to a number of \textit{sum rules} for quantized transport signatures of such topological edge modes in periodically driven systems~\cite{zeropeak2a}. In particular, it has been argued that the total sum of conductances measured for a system with FMFs, when chemical potentials is set to all even or odd (depending on the flavor of the FMF) multiples of half of the drive's frequency, is quantized. This Floquet sum rule is the generalization of the zero-bias conductance peak of the static Majorana bound states~\cite{zeropeak2a}. 
	
	In this paper, we study a Josephson junction of two driven superconductors which host FMF as edge states. Whereas similar setups have been studied in Ref.~\cite{Majref6,Majref6a}, it remains unclear whether they also give rise to 4$\pi$-periodic Josephson signature, and, if they do, whether this can be understood in terms of their steady-state occupation probabilities. In the case of static Majorana edge states, the 4$\pi$-periodic nature is a result of two Majorana edge states at the junction becoming degenerate and exchanging occupations from fully occupied to unoccupied when the phase difference across the junction is tuned through $\pi$~\cite{fracjos1,fracjos2,fracjos3}. In a setup where fermion parity is not conserved, this results in a sharp jump of Josephson current at $\pi$ phase difference. As we describe in the paper, we find that FMF can give rise to a similar signature. We also investigate the conditions under which a sharp jump of Josephson current at $\pi$ phase difference is obtained, reflecting the non-equilibrium nature of the system. Furthermore, we formulate the occupation of steady states in a Floquet system which is weakly connected to a thermal environment. This leads to the analytical understanding of the unusual Josephson signature of FMF in terms of their occupation. Using this formalism, we uncover a general Josephson Floquet sum rule, which extends the sum rules previously discussed for transport in periodically driven systems. In case of Josephson current, the sum rule is an exact counterpart of the signature of static Majorana edge states. Finally, we test the robustness of these unusual signatures of FMF in presence of static random impurities in the system.
	
	\section{Steady-state Floquet Josephson current}%
	We consider a Josephson junction between two driven superconductors hosting Floquet Majorana edge states with a phase difference $\phi = \phi_1-\phi_2$ and tunneling amplitude $w_J$ across the junction. Since the entire system is driven periodically with a Hamiltonian $h(t) = h(t+T)$, the solutions to the Sch\"odinger equation are $\ket{\psi_\alpha(t)} = e^{-i\epsilon_\alpha t} \ket{u_{\alpha}(t)}$, where $\epsilon_\alpha$ and $\ket{u_{\alpha}(t+T)} = \ket{u_{\alpha}(t)}$ are quasienergies and Floquet states, respectively, satisfying the eigenvalue equation
	$\left[ h(t) - i\partial_t \right] \ket{u_{\alpha}(t)} = \epsilon_\alpha \ket{u_{\alpha}(t)}$. (We set $\hbar=1$.) We take the quasienergies to be in the first Floquet zone $|\epsilon_\alpha| \leq \pi/T$ and identify quasienergies $\pm\pi/T$. Since the (mean-field) quasienergy spectrum has the usual particle-hole symmetry, FMFs can exist at quasienergies $\epsilon_b = b/T$ with $b=0,\pi$, often referred to as 0 and $\pi$ FMFs, respectively. 
	
	\begin{figure}[t]
		\centering
		\includegraphics[width=0.95\linewidth]{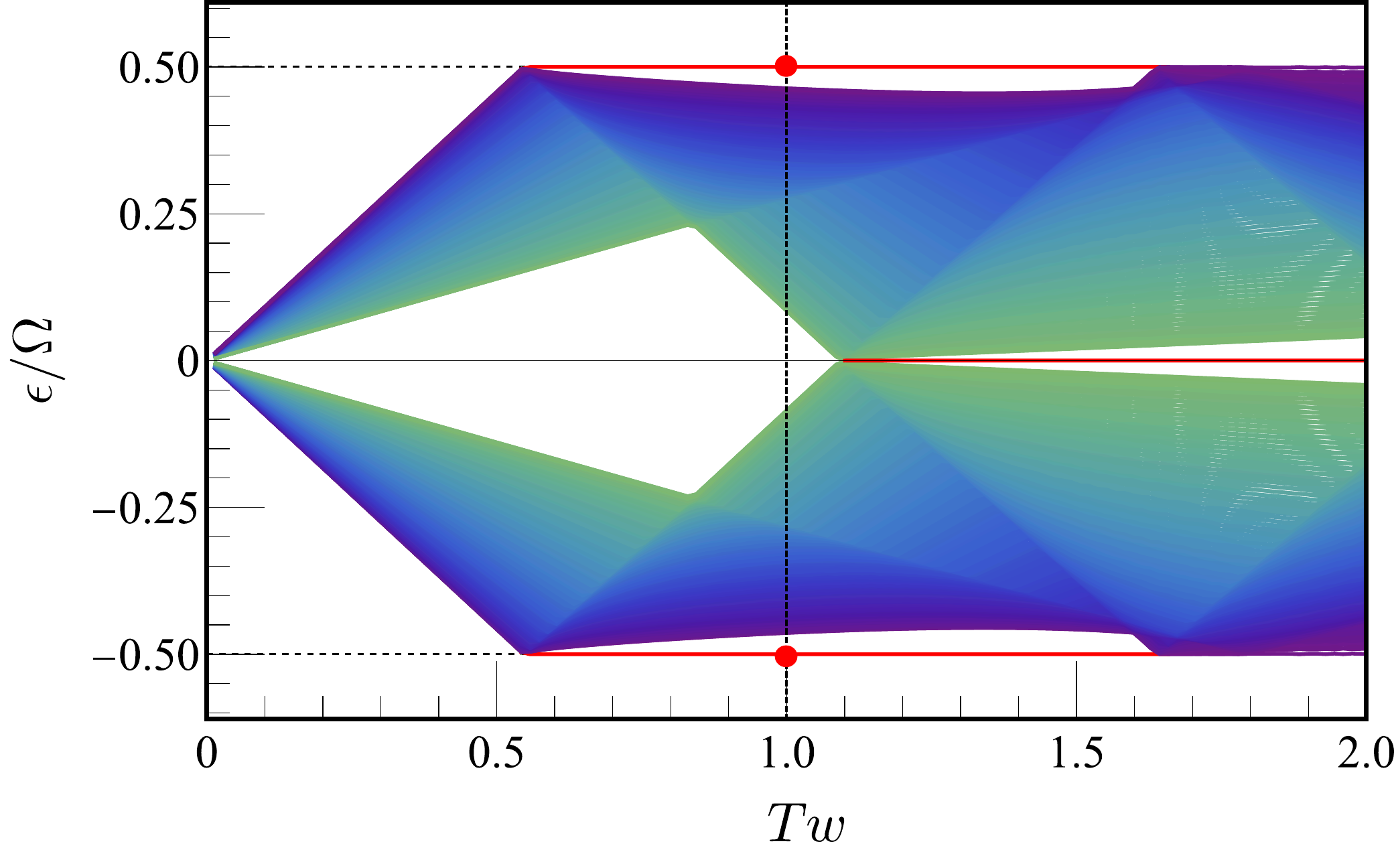}
		\caption{Quasienergy spectrum of a single driven Kitaev chain, Eq.~(\ref{eq:HSC}), with $N=200$ sites. Here, $\Delta/w=0.5$, $\mu_{0}/w=3.75$, $\mu_{1}/w=1.25$. For $T=1/w$, there are two $\pi$-FMFs}
		\label{fig:qes}
	\end{figure}
	
	We are interested in Josephson current flowing across the junction, $J(t) = \text{tr}[\rho(t) \partial_\phi h(t)]$, where $\rho(t)$ is the steady-state density matrix. In order to calculate the current, we assume the superconductors are weakly coupled to thermal leads at temperature $\theta_r$ and uniform chemical potential, $\mu_r$, i.e. at zero bias. 
	In this limit, the steady-state density matrix
	is approximately diagonal and time-independent in the Floquet basis, i.e. $\bra{u_\alpha(t)}{\rho(t)}\ket{u_\beta(t)} \approx n_{\alpha}(\mu_r)\delta_{\alpha\beta}$, with populations
	\begin{align}\label{eq:occu}
		n_{\alpha}(\mu_r)
		=\sum_{k\in\mathbb{Z}} f_r(\epsilon_{\alpha}+k\Omega-\mu_r) \braket{u^{(k)}_{\alpha}|u^{(k)}_{\alpha}},
	\end{align}
	where $f_r(x) = (1+e^{x/\theta_r})^{-1}$ is the reservoir's Fermi function, the Fourier modes $\ket{u^{(k)}_{\alpha}} = \int_0^T e^{ik\Omega t}|u_{\alpha}(t)\ra dt/T$, and $\Omega = 2\pi/T$ is the drive frequency. We provide the details of the derivation at Appendix~B. Such as occupation distribution was also previously discussed in Ref.~\onlinecite{occu_PhysRevX}.
	This is in fact true for the steady state of any periodically driven quantum system weakly coupled to a thermal fermion reservoir and can be understood as the incoherent mixture of Floquet sidebands indexed by $k$.
	
	We proceed with the Josephson current, which is written as $J(t)
	=\sum_{\alpha}n_{\alpha}(\mu_r) \big[\partial_{\phi}\epsilon_{\al} + i\partial_t \braket{u_{\alpha}(t)|\partial_{\phi}u_{\alpha}(t)}\big].$ 
	Using the particle-hole symmetry in Eq.~\eqref{eq:occu}, we can write the time-average 
	\begin{align}\label{eq:Jsumal}
		\bar J(\mu_r) =\frac1T\int_0^T J(t) dt = \sum_{\epsilon_{\alpha}<0}\nu_{\alpha} (\mu_r)\partial_{\phi}\epsilon_{\al}.
	\end{align}
	Here $\nu_{\alpha}(\mu_r) = n_{\alpha}(\mu_r) - n_{\bar{\alpha}}(\mu_r)$, where $\alpha,\bar{\alpha}$ states are related by particle-hole operation, i.e, $\epsilon_{\bar{\alpha}} = -\epsilon_{\alpha}$.
	This simplified expression, as we show below, captures the Josephson current with high accuracy. This is our first main result.
	
	\begin{figure}[t]
		\centering
		\includegraphics[width=0.90\linewidth]{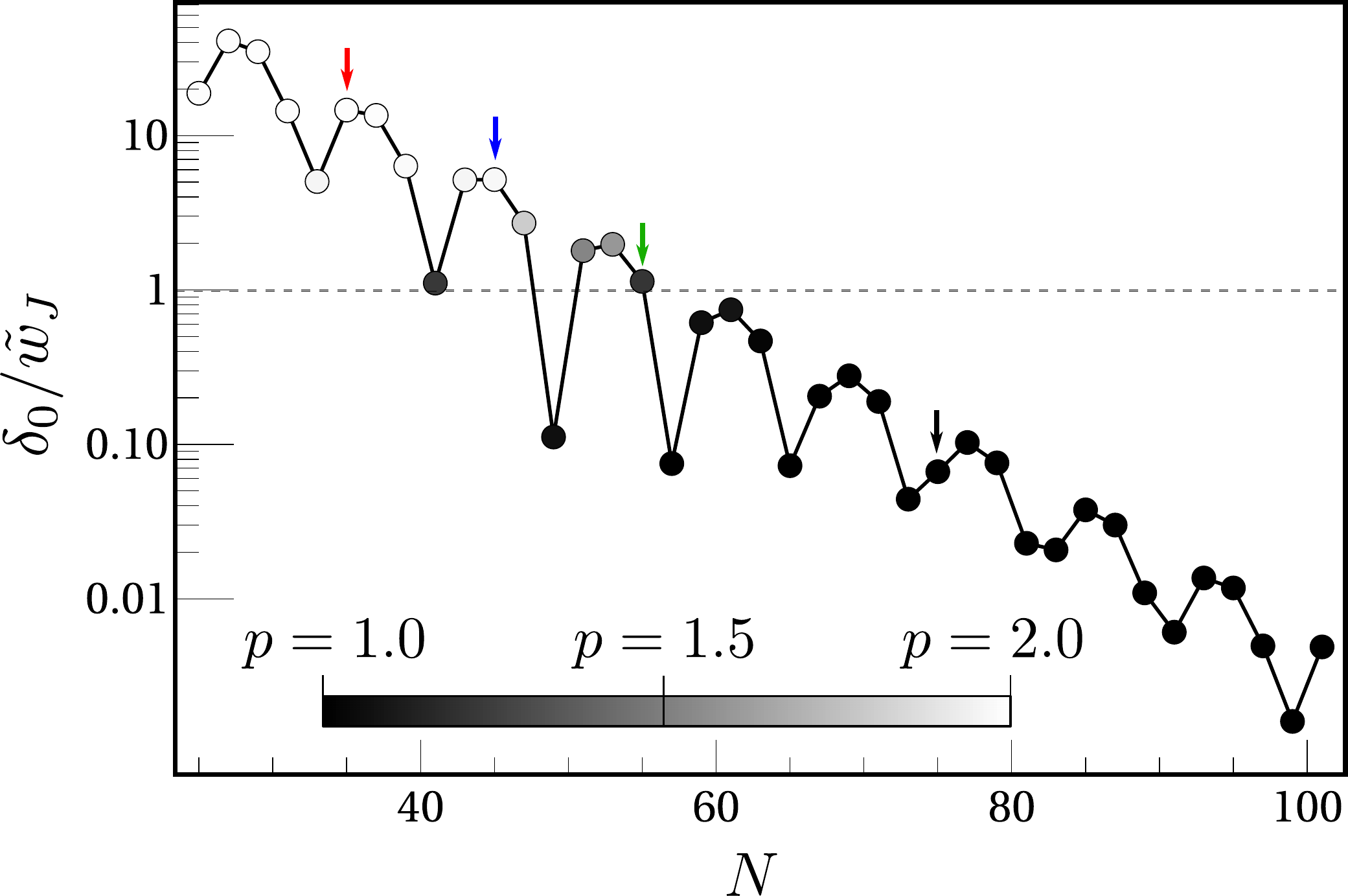}
		\caption{Quasienergy splitting of FMFs at the junction between two driven Kitaev chains as a function of the length of the chain ($N$ number of sites in the single superconductor). Here, $w_J/w=10^{-3}$ and $\tilde w_J = 2\sum_{k}|u_b^{(k)}|_J^2 w_J$ is normalized by the weight of the FMFs at the junction. The grayscale shows the exponent $p:=d\ln \bar J/d\ln w_J$, where the numerical computation of current is based on NEFG techniques (see the main text), at zero temperature. $T=1/w$ and other parameters are the same as in Fig.~\ref{fig:qes}}
		\label{fig:gaps}
	\end{figure}
	
	\section{Josephson current signatures of FMFs}%
	Let us first briefly recall the properties of the Josephson current in the presence of \emph{static} Majorana fermions. Projecting to the two-level system formed by the Majorana bound states, the tunneling between the superconductors splits the zero-energy states at the junction to $\pm E_J$ with $E_J(\phi) = \sqrt{\tilde w_J^2\cos^2(\phi/2) + E_0^2}$, where $E_0$ is the energy splitting of the Majorana bound states in the absence of the junction ($w_J=0$), for instance, due to finite size, and $\tilde w_J \propto w_J$ with a factor of the Majorana bound state wave functions at the junction. Consequently, the current 
	$
	J \approx \partial_\phi E_J = (\tilde w_J/2)^2 \sin\phi /E_J(\phi).
	$
	Thus, when $w_J \ll E_0$, the current $J\propto (\tilde{w}_J/2 E_0)^2\sin\phi$ is $2\pi$-periodic in $\phi$. On the other hand, when $w_J \gg E_0$, e.g. for large system sizes, the current $J\propto \eta \tilde{w}_J \sin(\phi/2)$ where $\eta = \pm$ is the fermion parity determined by the occupation of the split levels. The linear dependence on $w_J$ and $4\pi$-periodicity in $\phi$ for conserved $\eta$ or, alternatively, the finite jumps associated with switching fermion parity $\eta$ are telltale signatures of Majorana bound states.
	
	\begin{figure*}[t]
		\includegraphics[width=0.99\linewidth]{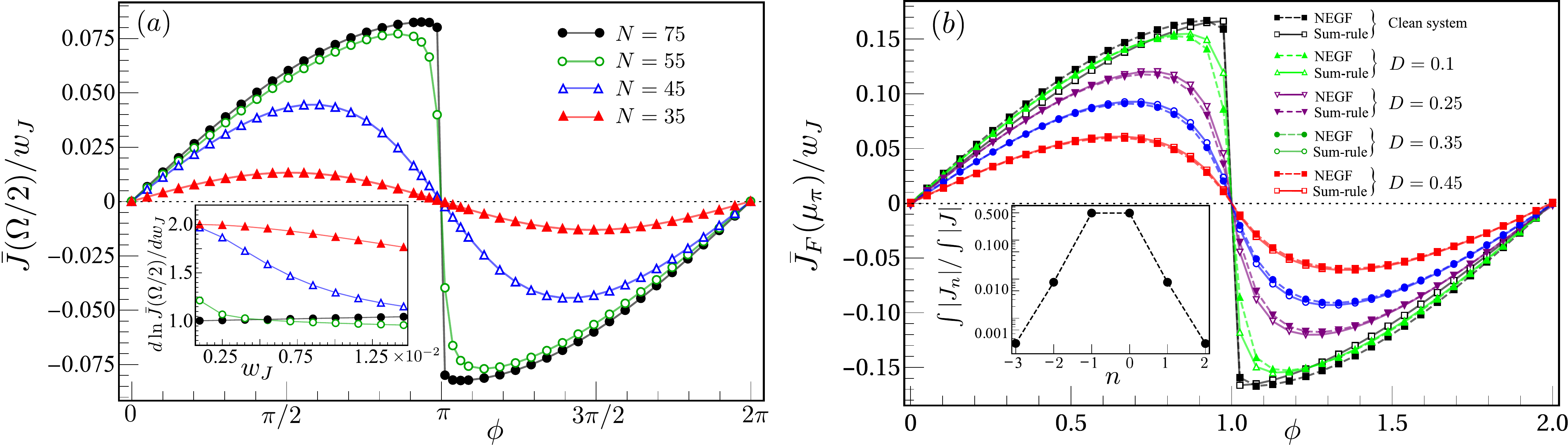}
		\caption{(a) Current-phase relationship for the Josephson junction between two driven Kitaev chains hosting $\pi$-FMFs for $\mu_{r=\pi}=\pi/T$. As the length of the chains, $N$, increases the time-averaged Josephson current $\bar{ J}(\Omega/2)$ crosses over from a smooth $w_J^2 \sin\phi$ form to $\eta w_J \sin(\phi/2)$ accompanied by jumps when the fermion parity $\eta$ switches sign. The inset shows the evolution of the exponent $p:=d\ln\bar J(\Omega/2)/ d\ln w_J$ as a function of $w_J$. (b) The sum rule for $\bar J_F(\mu_{b=\pi}) = \sum_{m\in\mathbb{Z}} J(\mu_b+m\Omega)$ and its robustness in presence of static disorder characterized by disorder strength, $D$ in units of $\Delta$. The analytical calculation (dashed), obtained using Eq.~\eqref{eq:Jsum} agrees very well with the numerical calculation (solid), obtained using non-equilibrium Green's functions. The inset shows the contributions from different harmonics quantified as $j_m = \oint|\bar J(\mu_b + m\Omega)|d\phi / \oint|\bar J_F|d\phi$. The biggest contribution comes from $m=-1,0$. In both panels, $T=1/w$ and other parameters are the same as in Fig.~\ref{fig:qes}.}
		\label{fig:jvsph}
	\end{figure*}
	
	Now, we note that when a single driven superconductor hosts FMFs at quasienergy $\epsilon_b$, the coupled system hosts four FMFs, two at faraway boundaries with quasienergies $\pm(\epsilon_b+e^{ib}\delta)$ and the other two at the junction with quasienergies $\pm(\epsilon_b+e^{ib}\delta_J) 
	$. In particular, we have $\delta_J \approx \sqrt{\tilde w_J^2\cos^2(\phi/2)+\delta_0^2}$, where $\delta_0$ is the FMFs quasienergy splitting without the junction ($w_J=0$) and $\tilde w_J \approx 2\sum_k|u_{b}^{(k)}|_J^2w_J$, with $|u_{b}^{(k)}|_J^2$ being the amplitude of the $k$th Fourier mode of the FMFs at the junction site (denoted by $J$).
	
	Thus, using Eq.~\eqref{eq:Jsumal}, we see that at sufficiently low temperature and high frequency, the phenomenology of the Josephson current is similar to the static case. The Josephson current can be expressed as:
	\begin{equation}
		\bar J \propto (\eta w_J)^p\sin(p\phi/2),
	\end{equation}
	shows a crossover at $\tilde w_J \sim \delta_0$, tunable by the system size, from $p=2$ for $w_J \ll \delta_0$ to $p=1$ for $w_J \gg \delta_0$. 
	\vspace{2mm}
	\emph{Josephson Floquet sum rule : }%
	A consequence of the above is a {sum rule} for the Josephson current over values of the chemical potential varied by the drive harmonics, i.e.
	\begin{equation}
		\bar J_F(\mu_r) := \sum_{m\in\mathbb{Z}} \bar J(\mu_r+m\Omega) \equiv \bar J_F(\mu_r+\Omega).
	\end{equation}
	one can restrict $|\mu_r|<\Omega/2$. Now, even though the summation is over all the negative quasi-energy states in Eq.~(\ref{eq:Jsumal}), as long as there occupation difference, $\nu_{b}$, of FMFs at the junction is finite, their contribution dominates. In this limit one can show that, at small temperature, such that $\Omega\theta_r\gg 1$ (for details, see Appendix~B),
	\begin{align}
		\nu_{b}^{\rm F} := \sum_{m\in\mathbb{Z}} \nu_{b}(\mu_{b} + m\Omega) \approx &~e^{ib}{\tanh}\left(\frac{\delta_J}{2\theta_r}\right) -e^{ib} \frac{2\delta_J}{\Omega},
	\end{align}
	where $\mu_b =- b/T$. At zero temperature and for large system size the above reduces to $\nu_{0/\pi}^{\rm F} =\pm1$, which is a \textit{sum rule} of the FMF occupation. With these considerations, one arrives at a rather simple expression of the summed current in the presence of FMFs, 
	\begin{align}\label{eq:Jsum}
		\bar{J}_F(\mu_b) \approx \nu_{b}^{\rm F} \frac{\partial \delta_{J}}{\partial\phi},
	\end{align}
	where $i$s are all the $b$ type FMFs modes, with $\mu_{b}$ is set at $b/T$. This is our second main result.
	
	\section{Lattice model and numerical simulation}%
	FMFs appear in the driven Kitaev chain~\cite{Majref6,Majref6a,MajsigAK,zeropeak2}, a driven one-dimensional $p$-wave superconductor with the Hamiltonian $\hat H(t) = \sum_{r,s=1}^{N}\hat \Psi^{\dagger}_{r} h_{rs}(t) \hat\Psi^\nodag_{s}$, where the Nambu spinor $\hat\Psi^{\dagger}_{r}=\left(e^{-i\phi/2} \hat c^{\dagger}_{r},e^{i\phi/2} \hat c^\nodag_{r}\right)$ with $\hat c^{\dagger}_{r}$ the fermion creation operator at site $r$, and
	\begin{align}\label{eq:HSC}
		h_{rs}(t) = \delta_{r\pm 1,s} (w\tau_z \pm i\Delta\tau_y)- 2\delta_{r,s}\mu(t) \tau_z .
	\end{align}
	Here, Pauli matrices $\tau_x, \tau_y, \tau_z$ act on the Nambu space, $\mu$ is the chemical potential, $w$ is the nearest neighbor hopping amplitude, and $\Delta$ and $\phi$ are, respectively, the amplitude and phase of the superconducting order parameter. We impose open boundary conditions by dropping terms with $r,s < 1$ and $>N$. 
	We take a two-step periodic drive
	\begin{equation}
		\mu(t)=\mu(t+T)=\begin{cases}
			\mu_0+\mu_1&0< t\leq T/2, \\
			\mu_0-\mu_1&T/2 < t\leq T.
		\end{cases}
	\end{equation}
		
	\begin{figure}[th]
		\centering
		\includegraphics[width=90mm,scale=0.5]{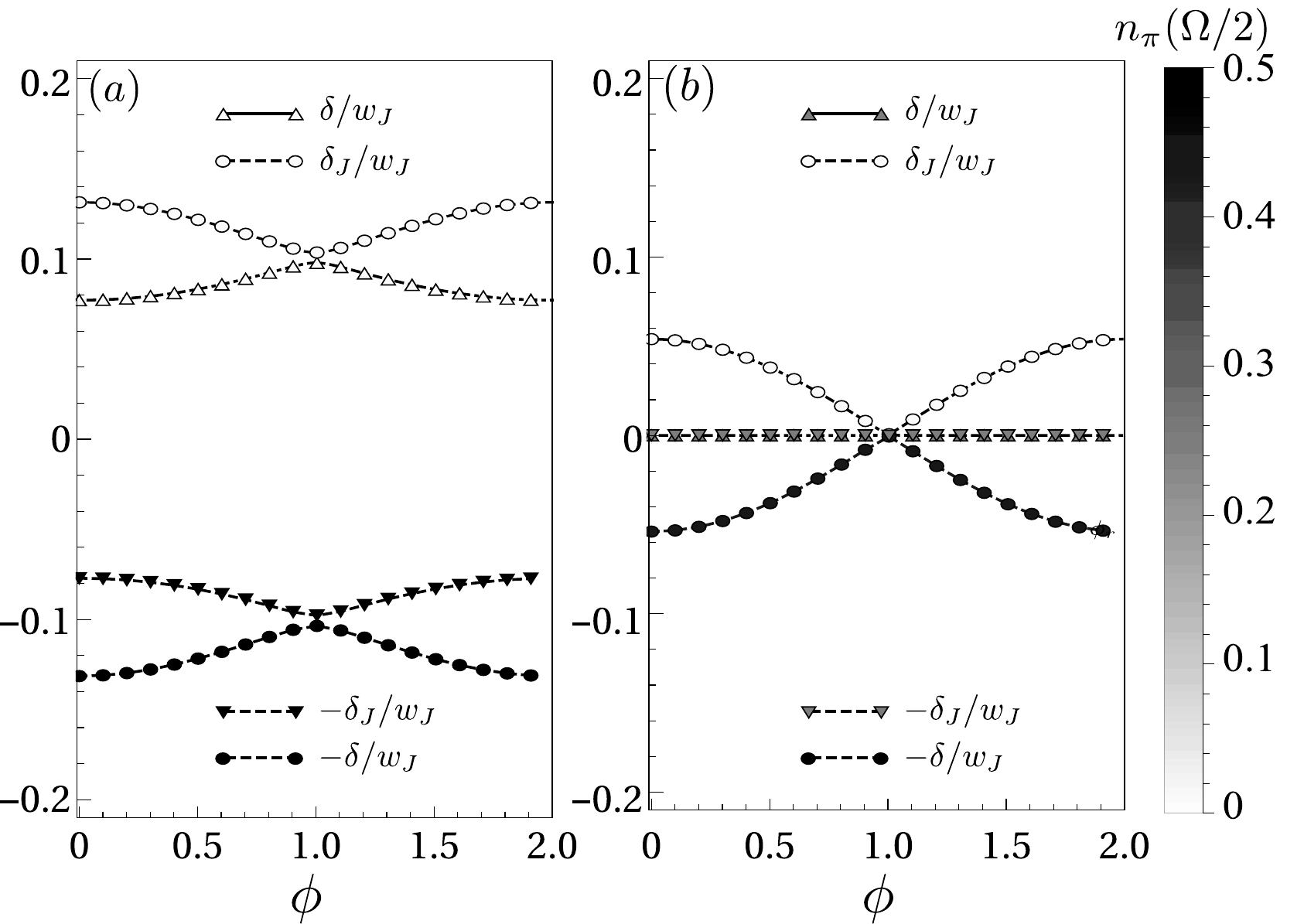}
		\caption{Changes in the quasi-energies and occupation probabilities of $\pi-$FMFs as a function of phase difference for system sizes 25 in (a) and 65 in (b). In the second case for large systems, FMFs localized at the ends of two superconductors do not contribute to the current and current changes sharply at $\phi=\pi$, which shows that it is $4\pi$ periodic. ($\mu_r=\Omega/2$)}
		\label{fig:E_occu}
	\end{figure}
	
	For a typical choice of parameters, the quasienergy spectrum of an open driven Kitaev chain is shown in Fig.~\ref{fig:qes}, highlighting the FMFs at $\epsilon_0$ and $\epsilon_\pi$. 
	For a junction between two chains, the Hamiltonian is $h_1(t) + h_2(t) + h_J$, where 
	\begin{equation}
		h_J = 
		w_J \tau_z e^{-i\phi \tau_z/2}, 
	\end{equation}
	is the junction Hamiltonian. 
	The quasienergy splittings $\delta$ and $\delta_J$ depend on system parameters, in particular the length $N$ of the chains and the phase difference $\phi = \phi_1-\phi_2$. For $w_J \ll \delta_0$, the spectra are largely unperturbed and $\delta\approx \delta_J \approx \delta_0 \propto e^{-N/\xi}$, with $\xi$ a localization length. On the other hand, for $w_J \gtrsim \delta_0$, the composite system essentially behaves as a single chain with length $2N$, thus $\delta \propto e^{-2N/\xi}$ and $\delta_J = \sqrt{\tilde w_J^2\cos^2 \phi + \delta_0^2}$ depends strongly on $\phi$. 
	
	\vspace{2mm}	
\section{Discussion}
	We calculate the current using both the analytical expression, Eq.~\eqref{eq:Jsumal}, and numerically using Floquet Green's functions (for details, see Appendix~C). In Fig.~(\ref{fig:gaps}), we show the FMFs quasienergy splitting, $\delta_0$, as well as the power-law scaling of the time-averaged Josephson current $\bar J(\Omega/2) \propto w_J^p$ as a function of system size. This clearly illustrates the crossover from quadratic ($p=2$) to linear ($p=1$) behavior at $\tilde w_J \sim \delta_0$. For sufficiently large $N$, the quasienergy splittings can also be estimated using Floquet perturbation theory (see Appendix D).
	
	In Fig.~(\ref{fig:jvsph}) we show the current-phase relationship for the Josephson current and its Floquet sum rule using numerically calculated Floquet Green's functions and compare them to analytical calculations for a junction hosting $\pi$-FMFs. As illustrated in Fig.~\ref{fig:jvsph}(a), the Josephson current at chemical potential $\mu_r=\Omega/2$ crosses over, with increasing size of the superconducting chains, from a smooth quadratic form $\propto w_J^2\sin\phi$ to a linear form $\propto\eta w_J\sin(\phi/2)$ that exhibits jumps at $\phi=\pi$ associated with switches in the fermion parity $\eta$ of $\pi$-FMFs. The same behavior is observed for all $\mu_r = m\Omega/2$, when $m$ is an odd integer. A similar phenomenology arises for junctions hosting $0$-FMFs and $\mu_r = m\Omega/2$, when $m$ is an even integer.
	
	The crossover regime matches well the point at which $\pi$-FMFs quasienergy splitting ratio $\delta_0/w_J \sim 1$ (see Fig.~(\ref{fig:gaps})). This behavior is captured very well by our simple analytical expressions of Josephson current, Eq.~(\ref{eq:Jsumal}), and its sum rule, Eq.~\eqref{eq:Jsum}. This is best seen in the Floquet sum rule for Josephson current, illustrated in Fig.~\ref{fig:jvsph}(b). However, we note that, as shown in the inset of Fig.~\ref{fig:jvsph}(b), for the parameter we consider the primary contributions to the sum rule come from $\mu_r = \pm\Omega/2$. 
	
	We also observe that, for large $N$ and small $w_J$, the current is carried almost entirely by the FMFs and the bulk contribution to the current is negligible. This can be understood as yet another consequence of the linear dependence of FMFs contributions on $w_J$ versus the quadratic dependence of bulk contributions. 
	This observation extends the explanation of sharp jumps in Josephson current from static to Floquet systems: when $\delta_0 \ll w_J$, the two FMFs at the junction are exchanged at the time-reversal symmetric point $\phi = \pi$; when the two FMFs at the junction have a difference in their population, this exchange leads to a sharp change in Josephson current. This is further explained in the Fig.~(\ref{fig:E_occu}), where we consider a smaller and a larger system size, showing notably different behaviors and occupation probabilities of the FMFs. For a larger system size (in Fig.~\ref{fig:E_occu}(b)), we observe that the $\pi$-FMFs at the junction with quasienegies $\pm \Omega/2\mp \delta_J$ exchange their occupation probability at $\phi=\pi$ and this results in a jump of the Josephson current. In this respect, the chemical potential $\mu_r$ of the reservoirs plays an important role: setting $\mu_r = m\Omega + b/T$ for a junction hosting $b$-FMFs gives rise to such a difference in occupation and jumps in the current, which are otherwise lost.
	
	Due to their topological origin, one expects the Josephson-current signatures of FMFs to be robust against perturbations and disorder.
	We study this robustness in the presence of static disorder, modeled by replacing the static chemical potential in the chains $\mu_0 \rightarrow \mu_0 + \delta\mu_i$, where $\delta\mu_i$ are taken randomly from an uncorrelated normal distribution of standard deviation $D$, which characterizes the strength of the disorder. As shown in Fig.~\ref{fig:jvsph}(b), the sharp jump in the Floquet sum rule is robust against weak disorder. However, the analytical expression, Eq.~\eqref{eq:Jsum}, remains valid across the entire range of disorder strengths reported here.
	
	Additional note: during preparation of the manuscript, another work Ref.~\onlinecite{matsyshyn2023fermi} has been posted which discusses occupation of Floquet states in agreement with our result,  Eq.~(\ref{eq:occu}) and Appendix B.
	
	\vspace{2mm}
	\emph{Acknowledgments-} A.K acknowledges support from the SERB (Govt. of India) via saction no. ECR/2018/00 1443, DAE (Govt. of India) via sanction no. 58/20/ 15/2019-BRNS, as well as MHRD (Govt. of India) via sanction no. SPARC/2018-2019/P538/SL. B.S. acknowledges support from the DOE grant DE-SC0020343, the College of Arts and Sciences and the Vice Provost for Research at Indiana University, Bloomington through the Faculty Research Support Program. R. K. acknowledges the use of PARAM Sanganak  and HPC facility at IIT Kanpur. The support and the resources provided by PARAM Sanganak under the National Supercomputing Mission, Government of India at the Indian Institute of Technology, Kanpur are gratefully acknowledged.
	
	\bibliography{ref2.bib}		

\begin{thebibliography}{80}%
\makeatletter
\providecommand \@ifxundefined [1]{%
 \@ifx{#1\undefined}
}%
\providecommand \@ifnum [1]{%
 \ifnum #1\expandafter \@firstoftwo
 \else \expandafter \@secondoftwo
 \fi
}%
\providecommand \@ifx [1]{%
 \ifx #1\expandafter \@firstoftwo
 \else \expandafter \@secondoftwo
 \fi
}%
\providecommand \natexlab [1]{#1}%
\providecommand \enquote  [1]{``#1''}%
\providecommand \bibnamefont  [1]{#1}%
\providecommand \bibfnamefont [1]{#1}%
\providecommand \citenamefont [1]{#1}%
\providecommand \href@noop [0]{\@secondoftwo}%
\providecommand \href [0]{\begingroup \@sanitize@url \@href}%
\providecommand \@href[1]{\@@startlink{#1}\@@href}%
\providecommand \@@href[1]{\endgroup#1\@@endlink}%
\providecommand \@sanitize@url [0]{\catcode `\\12\catcode `\$12\catcode
  `\&12\catcode `\#12\catcode `\^12\catcode `\_12\catcode `\%12\relax}%
\providecommand \@@startlink[1]{}%
\providecommand \@@endlink[0]{}%
\providecommand \url  [0]{\begingroup\@sanitize@url \@url }%
\providecommand \@url [1]{\endgroup\@href {#1}{\urlprefix }}%
\providecommand \urlprefix  [0]{URL }%
\providecommand \Eprint [0]{\href }%
\providecommand \doibase [0]{https://doi.org/}%
\providecommand \selectlanguage [0]{\@gobble}%
\providecommand \bibinfo  [0]{\@secondoftwo}%
\providecommand \bibfield  [0]{\@secondoftwo}%
\providecommand \translation [1]{[#1]}%
\providecommand \BibitemOpen [0]{}%
\providecommand \bibitemStop [0]{}%
\providecommand \bibitemNoStop [0]{.\EOS\space}%
\providecommand \EOS [0]{\spacefactor3000\relax}%
\providecommand \BibitemShut  [1]{\csname bibitem#1\endcsname}%
\let\auto@bib@innerbib\@empty
\bibitem [{\citenamefont {Majorana}(1937)}]{Ettore_1937}%
  \BibitemOpen
  \bibfield  {author} {\bibinfo {author} {\bibfnamefont {E.}~\bibnamefont
  {Majorana}},\ }\bibfield  {title} {\bibinfo {title} {Unpaired majorana
  fermions in quantum wires},\ }\href {https://doi.org/10.1007/BF02961314}
  {\bibfield  {journal} {\bibinfo  {journal} {Nuovo Cimento}\ }\textbf
  {\bibinfo {volume} {14}},\ \bibinfo {pages} {171} (\bibinfo {year}
  {1937})}\BibitemShut {NoStop}%
\bibitem [{\citenamefont {Kitaev}(2001)}]{Kitaev_2001}%
  \BibitemOpen
  \bibfield  {author} {\bibinfo {author} {\bibfnamefont {A.~Y.}\ \bibnamefont
  {Kitaev}},\ }\bibfield  {title} {\bibinfo {title} {Unpaired majorana fermions
  in quantum wires},\ }\href {https://doi.org/10.1070/1063-7869/44/10s/s29}
  {\bibfield  {journal} {\bibinfo  {journal} {Physics-Uspekhi}\ }\textbf
  {\bibinfo {volume} {44}},\ \bibinfo {pages} {131} (\bibinfo {year}
  {2001})}\BibitemShut {NoStop}%
\bibitem [{\citenamefont {Alicea}(2012)}]{Alicea_2012}%
  \BibitemOpen
  \bibfield  {author} {\bibinfo {author} {\bibfnamefont {J.}~\bibnamefont
  {Alicea}},\ }\bibfield  {title} {\bibinfo {title} {New directions in the
  pursuit of majorana fermions in solid state systems},\ }\href
  {https://doi.org/10.1088/0034-4885/75/7/076501} {\bibfield  {journal}
  {\bibinfo  {journal} {Reports on Progress in Physics}\ }\textbf {\bibinfo
  {volume} {75}},\ \bibinfo {pages} {076501} (\bibinfo {year}
  {2012})}\BibitemShut {NoStop}%
\bibitem [{\citenamefont {Wimmer}\ \emph {et~al.}(2010)\citenamefont {Wimmer},
  \citenamefont {Akhmerov}, \citenamefont {Medvedyeva}, \citenamefont
  {Tworzyd\l{}o},\ and\ \citenamefont {Beenakker}}]{Wimmer_2010}%
  \BibitemOpen
  \bibfield  {author} {\bibinfo {author} {\bibfnamefont {M.}~\bibnamefont
  {Wimmer}}, \bibinfo {author} {\bibfnamefont {A.~R.}\ \bibnamefont
  {Akhmerov}}, \bibinfo {author} {\bibfnamefont {M.~V.}\ \bibnamefont
  {Medvedyeva}}, \bibinfo {author} {\bibfnamefont {J.}~\bibnamefont
  {Tworzyd\l{}o}},\ and\ \bibinfo {author} {\bibfnamefont {C.~W.~J.}\
  \bibnamefont {Beenakker}},\ }\bibfield  {title} {\bibinfo {title} {Majorana
  bound states without vortices in topological superconductors with
  electrostatic defects},\ }\href
  {https://doi.org/10.1103/PhysRevLett.105.046803} {\bibfield  {journal}
  {\bibinfo  {journal} {Phys. Rev. Lett.}\ }\textbf {\bibinfo {volume} {105}},\
  \bibinfo {pages} {046803} (\bibinfo {year} {2010})}\BibitemShut {NoStop}%
\bibitem [{\citenamefont {Hasan}\ and\ \citenamefont {Kane}(2010)}]{Kane_2010}%
  \BibitemOpen
  \bibfield  {author} {\bibinfo {author} {\bibfnamefont {M.~Z.}\ \bibnamefont
  {Hasan}}\ and\ \bibinfo {author} {\bibfnamefont {C.~L.}\ \bibnamefont
  {Kane}},\ }\bibfield  {title} {\bibinfo {title} {Colloquium: Topological
  insulators},\ }\href {https://doi.org/10.1103/RevModPhys.82.3045} {\bibfield
  {journal} {\bibinfo  {journal} {Rev. Mod. Phys.}\ }\textbf {\bibinfo {volume}
  {82}},\ \bibinfo {pages} {3045} (\bibinfo {year} {2010})}\BibitemShut
  {NoStop}%
\bibitem [{\citenamefont {Qi}\ and\ \citenamefont {Zhang}(2011)}]{Zhang_2011}%
  \BibitemOpen
  \bibfield  {author} {\bibinfo {author} {\bibfnamefont {X.-L.}\ \bibnamefont
  {Qi}}\ and\ \bibinfo {author} {\bibfnamefont {S.-C.}\ \bibnamefont {Zhang}},\
  }\bibfield  {title} {\bibinfo {title} {Topological insulators and
  superconductors},\ }\href {https://doi.org/10.1103/revmodphys.83.1057}
  {\bibfield  {journal} {\bibinfo  {journal} {Reviews of Modern Physics}\
  }\textbf {\bibinfo {volume} {83}},\ \bibinfo {pages} {1057} (\bibinfo {year}
  {2011})}\BibitemShut {NoStop}%
\bibitem [{\citenamefont {Linder}\ \emph {et~al.}(2010)\citenamefont {Linder},
  \citenamefont {Tanaka}, \citenamefont {Yokoyama}, \citenamefont {Sudb\o{}},\
  and\ \citenamefont {Nagaosa}}]{Linder_2010}%
  \BibitemOpen
  \bibfield  {author} {\bibinfo {author} {\bibfnamefont {J.}~\bibnamefont
  {Linder}}, \bibinfo {author} {\bibfnamefont {Y.}~\bibnamefont {Tanaka}},
  \bibinfo {author} {\bibfnamefont {T.}~\bibnamefont {Yokoyama}}, \bibinfo
  {author} {\bibfnamefont {A.}~\bibnamefont {Sudb\o{}}},\ and\ \bibinfo
  {author} {\bibfnamefont {N.}~\bibnamefont {Nagaosa}},\ }\bibfield  {title}
  {\bibinfo {title} {Unconventional superconductivity on a topological
  insulator},\ }\href {https://doi.org/10.1103/PhysRevLett.104.067001}
  {\bibfield  {journal} {\bibinfo  {journal} {Phys. Rev. Lett.}\ }\textbf
  {\bibinfo {volume} {104}},\ \bibinfo {pages} {067001} (\bibinfo {year}
  {2010})}\BibitemShut {NoStop}%
\bibitem [{\citenamefont {Leijnse}\ and\ \citenamefont
  {Flensberg}(2012)}]{Leijnse_2012}%
  \BibitemOpen
  \bibfield  {author} {\bibinfo {author} {\bibfnamefont {M.}~\bibnamefont
  {Leijnse}}\ and\ \bibinfo {author} {\bibfnamefont {K.}~\bibnamefont
  {Flensberg}},\ }\bibfield  {title} {\bibinfo {title} {Introduction to
  topological superconductivity and majorana fermions},\ }\href
  {https://doi.org/10.1088/0268-1242/27/12/124003} {\bibfield  {journal}
  {\bibinfo  {journal} {Semiconductor Science and Technology}\ }\textbf
  {\bibinfo {volume} {27}},\ \bibinfo {pages} {124003} (\bibinfo {year}
  {2012})}\BibitemShut {NoStop}%
\bibitem [{\citenamefont {Potter}\ and\ \citenamefont
  {Lee}(2010)}]{Potter_2010}%
  \BibitemOpen
  \bibfield  {author} {\bibinfo {author} {\bibfnamefont {A.~C.}\ \bibnamefont
  {Potter}}\ and\ \bibinfo {author} {\bibfnamefont {P.~A.}\ \bibnamefont
  {Lee}},\ }\bibfield  {title} {\bibinfo {title} {Multichannel generalization
  of kitaev's majorana end states and a practical route to realize them in thin
  films},\ }\href {https://doi.org/10.1103/PhysRevLett.105.227003} {\bibfield
  {journal} {\bibinfo  {journal} {Phys. Rev. Lett.}\ }\textbf {\bibinfo
  {volume} {105}},\ \bibinfo {pages} {227003} (\bibinfo {year}
  {2010})}\BibitemShut {NoStop}%
\bibitem [{\citenamefont {Lutchyn}\ \emph {et~al.}(2011)\citenamefont
  {Lutchyn}, \citenamefont {Stanescu},\ and\ \citenamefont
  {Sarma}}]{Lutchyn_2011}%
  \BibitemOpen
  \bibfield  {author} {\bibinfo {author} {\bibfnamefont {R.~M.}\ \bibnamefont
  {Lutchyn}}, \bibinfo {author} {\bibfnamefont {T.~D.}\ \bibnamefont
  {Stanescu}},\ and\ \bibinfo {author} {\bibfnamefont {S.~D.}\ \bibnamefont
  {Sarma}},\ }\bibfield  {title} {\bibinfo {title} {Search for majorana
  fermions in multiband semiconducting nanowires},\ }\bibfield  {journal}
  {\bibinfo  {journal} {Physical Review Letters}\ }\textbf {\bibinfo {volume}
  {106}},\ \href {https://doi.org/10.1103/physrevlett.106.127001}
  {10.1103/physrevlett.106.127001} (\bibinfo {year} {2011})\BibitemShut
  {NoStop}%
\bibitem [{\citenamefont {Beenakker}(2013)}]{Beenakker_2013}%
  \BibitemOpen
  \bibfield  {author} {\bibinfo {author} {\bibfnamefont {C.}~\bibnamefont
  {Beenakker}},\ }\bibfield  {title} {\bibinfo {title} {Search for majorana
  fermions in superconductors},\ }\href
  {https://doi.org/10.1146/annurev-conmatphys-030212-184337} {\bibfield
  {journal} {\bibinfo  {journal} {Annual Review of Condensed Matter Physics}\
  }\textbf {\bibinfo {volume} {4}},\ \bibinfo {pages} {113} (\bibinfo {year}
  {2013})},\ \Eprint
  {https://arxiv.org/abs/https://doi.org/10.1146/annurev-conmatphys-030212-184337}
  {https://doi.org/10.1146/annurev-conmatphys-030212-184337} \BibitemShut
  {NoStop}%
\bibitem [{\citenamefont {Stanescu}\ and\ \citenamefont
  {Tewari}(2013)}]{Stanescu_2013}%
  \BibitemOpen
  \bibfield  {author} {\bibinfo {author} {\bibfnamefont {T.~D.}\ \bibnamefont
  {Stanescu}}\ and\ \bibinfo {author} {\bibfnamefont {S.}~\bibnamefont
  {Tewari}},\ }\bibfield  {title} {\bibinfo {title} {Majorana fermions in
  semiconductor nanowires: fundamentals, modeling, and experiment},\ }\href
  {https://doi.org/10.1088/0953-8984/25/23/233201} {\bibfield  {journal}
  {\bibinfo  {journal} {Journal of Physics: Condensed Matter}\ }\textbf
  {\bibinfo {volume} {25}},\ \bibinfo {pages} {233201} (\bibinfo {year}
  {2013})}\BibitemShut {NoStop}%
\bibitem [{\citenamefont {Elliott}\ and\ \citenamefont
  {Franz}(2015)}]{MFranz_2015}%
  \BibitemOpen
  \bibfield  {author} {\bibinfo {author} {\bibfnamefont {S.~R.}\ \bibnamefont
  {Elliott}}\ and\ \bibinfo {author} {\bibfnamefont {M.}~\bibnamefont
  {Franz}},\ }\bibfield  {title} {\bibinfo {title} {Colloquium: Majorana
  fermions in nuclear, particle, and solid-state physics},\ }\href
  {https://doi.org/10.1103/RevModPhys.87.137} {\bibfield  {journal} {\bibinfo
  {journal} {Rev. Mod. Phys.}\ }\textbf {\bibinfo {volume} {87}},\ \bibinfo
  {pages} {137} (\bibinfo {year} {2015})}\BibitemShut {NoStop}%
\bibitem [{\citenamefont {Oreg}\ \emph {et~al.}(2010)\citenamefont {Oreg},
  \citenamefont {Refael},\ and\ \citenamefont {von Oppen}}]{Majref10}%
  \BibitemOpen
  \bibfield  {author} {\bibinfo {author} {\bibfnamefont {Y.}~\bibnamefont
  {Oreg}}, \bibinfo {author} {\bibfnamefont {G.}~\bibnamefont {Refael}},\ and\
  \bibinfo {author} {\bibfnamefont {F.}~\bibnamefont {von Oppen}},\ }\bibfield
  {title} {\bibinfo {title} {Helical liquids and majorana bound states in
  quantum wires},\ }\href {https://doi.org/10.1103/PhysRevLett.105.177002}
  {\bibfield  {journal} {\bibinfo  {journal} {Phys. Rev. Lett.}\ }\textbf
  {\bibinfo {volume} {105}},\ \bibinfo {pages} {177002} (\bibinfo {year}
  {2010})}\BibitemShut {NoStop}%
\bibitem [{\citenamefont {Sau}\ \emph {et~al.}(2010)\citenamefont {Sau},
  \citenamefont {Tewari},\ and\ \citenamefont {Das~Sarma}}]{Majref11}%
  \BibitemOpen
  \bibfield  {author} {\bibinfo {author} {\bibfnamefont {J.~D.}\ \bibnamefont
  {Sau}}, \bibinfo {author} {\bibfnamefont {S.}~\bibnamefont {Tewari}},\ and\
  \bibinfo {author} {\bibfnamefont {S.}~\bibnamefont {Das~Sarma}},\ }\bibfield
  {title} {\bibinfo {title} {Universal quantum computation in a semiconductor
  quantum wire network},\ }\href {https://doi.org/10.1103/PhysRevA.82.052322}
  {\bibfield  {journal} {\bibinfo  {journal} {Phys. Rev. A}\ }\textbf {\bibinfo
  {volume} {82}},\ \bibinfo {pages} {052322} (\bibinfo {year}
  {2010})}\BibitemShut {NoStop}%
\bibitem [{\citenamefont {Shivamoggi}\ \emph {et~al.}(2010)\citenamefont
  {Shivamoggi}, \citenamefont {Refael},\ and\ \citenamefont
  {Moore}}]{Majref12}%
  \BibitemOpen
  \bibfield  {author} {\bibinfo {author} {\bibfnamefont {V.}~\bibnamefont
  {Shivamoggi}}, \bibinfo {author} {\bibfnamefont {G.}~\bibnamefont {Refael}},\
  and\ \bibinfo {author} {\bibfnamefont {J.~E.}\ \bibnamefont {Moore}},\
  }\bibfield  {title} {\bibinfo {title} {Majorana fermion chain at the quantum
  spin hall edge},\ }\href {https://doi.org/10.1103/PhysRevB.82.041405}
  {\bibfield  {journal} {\bibinfo  {journal} {Phys. Rev. B}\ }\textbf {\bibinfo
  {volume} {82}},\ \bibinfo {pages} {041405} (\bibinfo {year}
  {2010})}\BibitemShut {NoStop}%
\bibitem [{\citenamefont {Neupert}\ \emph {et~al.}(2010)\citenamefont
  {Neupert}, \citenamefont {Onoda},\ and\ \citenamefont {Furusaki}}]{Majref13}%
  \BibitemOpen
  \bibfield  {author} {\bibinfo {author} {\bibfnamefont {T.}~\bibnamefont
  {Neupert}}, \bibinfo {author} {\bibfnamefont {S.}~\bibnamefont {Onoda}},\
  and\ \bibinfo {author} {\bibfnamefont {A.}~\bibnamefont {Furusaki}},\
  }\bibfield  {title} {\bibinfo {title} {Chain of majorana states from
  superconducting dirac fermions at a magnetic domain wall},\ }\href
  {https://doi.org/10.1103/PhysRevLett.105.206404} {\bibfield  {journal}
  {\bibinfo  {journal} {Phys. Rev. Lett.}\ }\textbf {\bibinfo {volume} {105}},\
  \bibinfo {pages} {206404} (\bibinfo {year} {2010})}\BibitemShut {NoStop}%
\bibitem [{\citenamefont {Deng}\ \emph {et~al.}(2012)\citenamefont {Deng},
  \citenamefont {Yu}, \citenamefont {Huang}, \citenamefont {Larsson},
  \citenamefont {Caroff},\ and\ \citenamefont {Xu}}]{zeropeak1exp}%
  \BibitemOpen
  \bibfield  {author} {\bibinfo {author} {\bibfnamefont {M.}~\bibnamefont
  {Deng}}, \bibinfo {author} {\bibfnamefont {C.~L.}\ \bibnamefont {Yu}},
  \bibinfo {author} {\bibfnamefont {G.~Y.}\ \bibnamefont {Huang}}, \bibinfo
  {author} {\bibfnamefont {M.}~\bibnamefont {Larsson}}, \bibinfo {author}
  {\bibfnamefont {P.}~\bibnamefont {Caroff}},\ and\ \bibinfo {author}
  {\bibfnamefont {H.~Q.}\ \bibnamefont {Xu}},\ }\bibfield  {title} {\bibinfo
  {title} {Anomalous zero-bias conductance peak in a nb-insb nanowire-nb hybrid
  device.},\ }\href@noop {} {\bibfield  {journal} {\bibinfo  {journal} {Nano
  letters}\ }\textbf {\bibinfo {volume} {12 12}},\ \bibinfo {pages} {6414}
  (\bibinfo {year} {2012})}\BibitemShut {NoStop}%
\bibitem [{\citenamefont {Finck}\ \emph {et~al.}(2013)\citenamefont {Finck},
  \citenamefont {Van~Harlingen}, \citenamefont {Mohseni}, \citenamefont
  {Jung},\ and\ \citenamefont {Li}}]{zeropeak2exp}%
  \BibitemOpen
  \bibfield  {author} {\bibinfo {author} {\bibfnamefont {A.~D.~K.}\
  \bibnamefont {Finck}}, \bibinfo {author} {\bibfnamefont {D.~J.}\ \bibnamefont
  {Van~Harlingen}}, \bibinfo {author} {\bibfnamefont {P.~K.}\ \bibnamefont
  {Mohseni}}, \bibinfo {author} {\bibfnamefont {K.}~\bibnamefont {Jung}},\ and\
  \bibinfo {author} {\bibfnamefont {X.}~\bibnamefont {Li}},\ }\bibfield
  {title} {\bibinfo {title} {Anomalous modulation of a zero-bias peak in a
  hybrid nanowire-superconductor device},\ }\href
  {https://doi.org/10.1103/PhysRevLett.110.126406} {\bibfield  {journal}
  {\bibinfo  {journal} {Phys. Rev. Lett.}\ }\textbf {\bibinfo {volume} {110}},\
  \bibinfo {pages} {126406} (\bibinfo {year} {2013})}\BibitemShut {NoStop}%
\bibitem [{\citenamefont {Mourik}\ \emph {et~al.}(2012)\citenamefont {Mourik},
  \citenamefont {Zuo}, \citenamefont {Frolov}, \citenamefont {Plissard},
  \citenamefont {Bakkers},\ and\ \citenamefont {Kouwenhoven}}]{zeropeak3exp}%
  \BibitemOpen
  \bibfield  {author} {\bibinfo {author} {\bibfnamefont {V.}~\bibnamefont
  {Mourik}}, \bibinfo {author} {\bibfnamefont {K.}~\bibnamefont {Zuo}},
  \bibinfo {author} {\bibfnamefont {S.~M.}\ \bibnamefont {Frolov}}, \bibinfo
  {author} {\bibfnamefont {S.~R.}\ \bibnamefont {Plissard}}, \bibinfo {author}
  {\bibfnamefont {E.~P. A.~M.}\ \bibnamefont {Bakkers}},\ and\ \bibinfo
  {author} {\bibfnamefont {L.~P.}\ \bibnamefont {Kouwenhoven}},\ }\bibfield
  {title} {\bibinfo {title} {Signatures of majorana fermions in hybrid
  superconductor-semiconductor nanowire devices},\ }\href
  {https://doi.org/10.1126/science.1222360} {\bibfield  {journal} {\bibinfo
  {journal} {Science}\ }\textbf {\bibinfo {volume} {336}},\ \bibinfo {pages}
  {1003} (\bibinfo {year} {2012})}\BibitemShut {NoStop}%
\bibitem [{\citenamefont {Churchill}\ \emph {et~al.}(2013)\citenamefont
  {Churchill}, \citenamefont {Fatemi}, \citenamefont {Grove-Rasmussen},
  \citenamefont {Deng}, \citenamefont {Caroff}, \citenamefont {Xu},\ and\
  \citenamefont {Marcus}}]{zeropeak4exp}%
  \BibitemOpen
  \bibfield  {author} {\bibinfo {author} {\bibfnamefont {H.~O.~H.}\
  \bibnamefont {Churchill}}, \bibinfo {author} {\bibfnamefont {V.}~\bibnamefont
  {Fatemi}}, \bibinfo {author} {\bibfnamefont {K.}~\bibnamefont
  {Grove-Rasmussen}}, \bibinfo {author} {\bibfnamefont {M.~T.}\ \bibnamefont
  {Deng}}, \bibinfo {author} {\bibfnamefont {P.}~\bibnamefont {Caroff}},
  \bibinfo {author} {\bibfnamefont {H.~Q.}\ \bibnamefont {Xu}},\ and\ \bibinfo
  {author} {\bibfnamefont {C.~M.}\ \bibnamefont {Marcus}},\ }\bibfield  {title}
  {\bibinfo {title} {Superconductor-nanowire devices from tunneling to the
  multichannel regime: Zero-bias oscillations and magnetoconductance
  crossover},\ }\bibfield  {journal} {\bibinfo  {journal} {Physical Review B}\
  }\textbf {\bibinfo {volume} {87}},\ \href
  {https://doi.org/10.1103/physrevb.87.241401} {10.1103/physrevb.87.241401}
  (\bibinfo {year} {2013})\BibitemShut {NoStop}%
\bibitem [{\citenamefont {Oostinga}\ \emph {et~al.}(2013)\citenamefont
  {Oostinga}, \citenamefont {Maier}, \citenamefont {Sch\"uffelgen},
  \citenamefont {Knott}, \citenamefont {Ames}, \citenamefont {Br\"une},
  \citenamefont {Tkachov}, \citenamefont {Buhmann},\ and\ \citenamefont
  {Molenkamp}}]{zeropeak5exp}%
  \BibitemOpen
  \bibfield  {author} {\bibinfo {author} {\bibfnamefont {J.~B.}\ \bibnamefont
  {Oostinga}}, \bibinfo {author} {\bibfnamefont {L.}~\bibnamefont {Maier}},
  \bibinfo {author} {\bibfnamefont {P.}~\bibnamefont {Sch\"uffelgen}}, \bibinfo
  {author} {\bibfnamefont {D.}~\bibnamefont {Knott}}, \bibinfo {author}
  {\bibfnamefont {C.}~\bibnamefont {Ames}}, \bibinfo {author} {\bibfnamefont
  {C.}~\bibnamefont {Br\"une}}, \bibinfo {author} {\bibfnamefont
  {G.}~\bibnamefont {Tkachov}}, \bibinfo {author} {\bibfnamefont
  {H.}~\bibnamefont {Buhmann}},\ and\ \bibinfo {author} {\bibfnamefont {L.~W.}\
  \bibnamefont {Molenkamp}},\ }\bibfield  {title} {\bibinfo {title} {Josephson
  supercurrent through the topological surface states of strained bulk hgte},\
  }\href {https://doi.org/10.1103/PhysRevX.3.021007} {\bibfield  {journal}
  {\bibinfo  {journal} {Phys. Rev. X}\ }\textbf {\bibinfo {volume} {3}},\
  \bibinfo {pages} {021007} (\bibinfo {year} {2013})}\BibitemShut {NoStop}%
\bibitem [{\citenamefont {Deacon}\ \emph {et~al.}(2017)\citenamefont {Deacon},
  \citenamefont {Wiedenmann}, \citenamefont {Bocquillon}, \citenamefont
  {Dom\'{\i}nguez}, \citenamefont {Klapwijk}, \citenamefont {Leubner},
  \citenamefont {Br\"une}, \citenamefont {Hankiewicz}, \citenamefont {Tarucha},
  \citenamefont {Ishibashi}, \citenamefont {Buhmann},\ and\ \citenamefont
  {Molenkamp}}]{fracjos1exp}%
  \BibitemOpen
  \bibfield  {author} {\bibinfo {author} {\bibfnamefont {R.~S.}\ \bibnamefont
  {Deacon}}, \bibinfo {author} {\bibfnamefont {J.}~\bibnamefont {Wiedenmann}},
  \bibinfo {author} {\bibfnamefont {E.}~\bibnamefont {Bocquillon}}, \bibinfo
  {author} {\bibfnamefont {F.}~\bibnamefont {Dom\'{\i}nguez}}, \bibinfo
  {author} {\bibfnamefont {T.~M.}\ \bibnamefont {Klapwijk}}, \bibinfo {author}
  {\bibfnamefont {P.}~\bibnamefont {Leubner}}, \bibinfo {author} {\bibfnamefont
  {C.}~\bibnamefont {Br\"une}}, \bibinfo {author} {\bibfnamefont {E.~M.}\
  \bibnamefont {Hankiewicz}}, \bibinfo {author} {\bibfnamefont
  {S.}~\bibnamefont {Tarucha}}, \bibinfo {author} {\bibfnamefont
  {K.}~\bibnamefont {Ishibashi}}, \bibinfo {author} {\bibfnamefont
  {H.}~\bibnamefont {Buhmann}},\ and\ \bibinfo {author} {\bibfnamefont {L.~W.}\
  \bibnamefont {Molenkamp}},\ }\bibfield  {title} {\bibinfo {title} {Josephson
  radiation from gapless andreev bound states in hgte-based topological
  junctions},\ }\href {https://doi.org/10.1103/PhysRevX.7.021011} {\bibfield
  {journal} {\bibinfo  {journal} {Phys. Rev. X}\ }\textbf {\bibinfo {volume}
  {7}},\ \bibinfo {pages} {021011} (\bibinfo {year} {2017})}\BibitemShut
  {NoStop}%
\bibitem [{\citenamefont {Lutchyn}\ \emph {et~al.}(2018)\citenamefont
  {Lutchyn}, \citenamefont {Bakkers}, \citenamefont {Kouwenhoven},
  \citenamefont {Krogstrup}, \citenamefont {Marcus},\ and\ \citenamefont
  {Oreg}}]{fracjos2exp}%
  \BibitemOpen
  \bibfield  {author} {\bibinfo {author} {\bibfnamefont {R.~M.}\ \bibnamefont
  {Lutchyn}}, \bibinfo {author} {\bibfnamefont {E.~P. A.~M.}\ \bibnamefont
  {Bakkers}}, \bibinfo {author} {\bibfnamefont {L.~P.}\ \bibnamefont
  {Kouwenhoven}}, \bibinfo {author} {\bibfnamefont {P.}~\bibnamefont
  {Krogstrup}}, \bibinfo {author} {\bibfnamefont {C.~M.}\ \bibnamefont
  {Marcus}},\ and\ \bibinfo {author} {\bibfnamefont {Y.}~\bibnamefont {Oreg}},\
  }\bibfield  {title} {\bibinfo {title} {Majorana zero modes in
  superconductor{\textendash}semiconductor heterostructures},\ }\href
  {https://doi.org/10.1038/s41578-018-0003-1} {\bibfield  {journal} {\bibinfo
  {journal} {Nature Reviews Materials}\ }\textbf {\bibinfo {volume} {3}},\
  \bibinfo {pages} {52} (\bibinfo {year} {2018})}\BibitemShut {NoStop}%
\bibitem [{\citenamefont {Laroche}\ \emph {et~al.}(2019)\citenamefont
  {Laroche}, \citenamefont {Bouman}, \citenamefont {van Woerkom}, \citenamefont
  {Proutski}, \citenamefont {Murthy}, \citenamefont {Pikulin}, \citenamefont
  {Nayak}, \citenamefont {Gulik}, \citenamefont {Nygård}, \citenamefont
  {Krogstrup}, \citenamefont {Kouwenhoven},\ and\ \citenamefont
  {Geresdi}}]{fracjos3exp}%
  \BibitemOpen
  \bibfield  {author} {\bibinfo {author} {\bibfnamefont {D.}~\bibnamefont
  {Laroche}}, \bibinfo {author} {\bibfnamefont {D.}~\bibnamefont {Bouman}},
  \bibinfo {author} {\bibfnamefont {D.}~\bibnamefont {van Woerkom}}, \bibinfo
  {author} {\bibfnamefont {A.}~\bibnamefont {Proutski}}, \bibinfo {author}
  {\bibfnamefont {C.}~\bibnamefont {Murthy}}, \bibinfo {author} {\bibfnamefont
  {D.}~\bibnamefont {Pikulin}}, \bibinfo {author} {\bibfnamefont
  {C.}~\bibnamefont {Nayak}}, \bibinfo {author} {\bibfnamefont
  {R.}~\bibnamefont {Gulik}}, \bibinfo {author} {\bibfnamefont
  {J.}~\bibnamefont {Nygård}}, \bibinfo {author} {\bibfnamefont
  {P.}~\bibnamefont {Krogstrup}}, \bibinfo {author} {\bibfnamefont
  {L.}~\bibnamefont {Kouwenhoven}},\ and\ \bibinfo {author} {\bibfnamefont
  {A.}~\bibnamefont {Geresdi}},\ }\bibfield  {title} {\bibinfo {title}
  {Observation of the 4pi-periodic josephson effect in indium arsenide
  nanowires},\ }\href {https://doi.org/10.1038/s41467-018-08161-2} {\bibfield
  {journal} {\bibinfo  {journal} {Nature Communications}\ }\textbf {\bibinfo
  {volume} {10}} (\bibinfo {year} {2019})}\BibitemShut {NoStop}%
\bibitem [{\citenamefont {Akhmerov}\ \emph {et~al.}(2011)\citenamefont
  {Akhmerov}, \citenamefont {Dahlhaus}, \citenamefont {Hassler}, \citenamefont
  {Wimmer},\ and\ \citenamefont {Beenakker}}]{zeropeak1}%
  \BibitemOpen
  \bibfield  {author} {\bibinfo {author} {\bibfnamefont {A.~R.}\ \bibnamefont
  {Akhmerov}}, \bibinfo {author} {\bibfnamefont {J.~P.}\ \bibnamefont
  {Dahlhaus}}, \bibinfo {author} {\bibfnamefont {F.}~\bibnamefont {Hassler}},
  \bibinfo {author} {\bibfnamefont {M.}~\bibnamefont {Wimmer}},\ and\ \bibinfo
  {author} {\bibfnamefont {C.~W.~J.}\ \bibnamefont {Beenakker}},\ }\bibfield
  {title} {\bibinfo {title} {Quantized conductance at the majorana phase
  transition in a disordered superconducting wire},\ }\href
  {https://doi.org/10.1103/PhysRevLett.106.057001} {\bibfield  {journal}
  {\bibinfo  {journal} {Phys. Rev. Lett.}\ }\textbf {\bibinfo {volume} {106}},\
  \bibinfo {pages} {057001} (\bibinfo {year} {2011})}\BibitemShut {NoStop}%
\bibitem [{\citenamefont {Law}\ \emph {et~al.}(2009)\citenamefont {Law},
  \citenamefont {Lee},\ and\ \citenamefont {Ng}}]{zeropeak2}%
  \BibitemOpen
  \bibfield  {author} {\bibinfo {author} {\bibfnamefont {K.~T.}\ \bibnamefont
  {Law}}, \bibinfo {author} {\bibfnamefont {P.~A.}\ \bibnamefont {Lee}},\ and\
  \bibinfo {author} {\bibfnamefont {T.~K.}\ \bibnamefont {Ng}},\ }\bibfield
  {title} {\bibinfo {title} {Majorana fermion induced resonant andreev
  reflection},\ }\href {https://doi.org/10.1103/PhysRevLett.103.237001}
  {\bibfield  {journal} {\bibinfo  {journal} {Phys. Rev. Lett.}\ }\textbf
  {\bibinfo {volume} {103}},\ \bibinfo {pages} {237001} (\bibinfo {year}
  {2009})}\BibitemShut {NoStop}%
\bibitem [{\citenamefont {Flensberg}(2010)}]{zeropeak3}%
  \BibitemOpen
  \bibfield  {author} {\bibinfo {author} {\bibfnamefont {K.}~\bibnamefont
  {Flensberg}},\ }\bibfield  {title} {\bibinfo {title} {Tunneling
  characteristics of a chain of majorana bound states},\ }\href
  {https://doi.org/10.1103/PhysRevB.82.180516} {\bibfield  {journal} {\bibinfo
  {journal} {Phys. Rev. B}\ }\textbf {\bibinfo {volume} {82}},\ \bibinfo
  {pages} {180516} (\bibinfo {year} {2010})}\BibitemShut {NoStop}%
\bibitem [{\citenamefont {Liu}\ \emph {et~al.}(2012)\citenamefont {Liu},
  \citenamefont {Potter}, \citenamefont {Law},\ and\ \citenamefont
  {Lee}}]{zeropeak4}%
  \BibitemOpen
  \bibfield  {author} {\bibinfo {author} {\bibfnamefont {J.}~\bibnamefont
  {Liu}}, \bibinfo {author} {\bibfnamefont {A.~C.}\ \bibnamefont {Potter}},
  \bibinfo {author} {\bibfnamefont {K.~T.}\ \bibnamefont {Law}},\ and\ \bibinfo
  {author} {\bibfnamefont {P.~A.}\ \bibnamefont {Lee}},\ }\bibfield  {title}
  {\bibinfo {title} {Zero-bias peaks in the tunneling conductance of
  spin-orbit-coupled superconducting wires with and without majorana
  end-states},\ }\href {https://doi.org/10.1103/PhysRevLett.109.267002}
  {\bibfield  {journal} {\bibinfo  {journal} {Phys. Rev. Lett.}\ }\textbf
  {\bibinfo {volume} {109}},\ \bibinfo {pages} {267002} (\bibinfo {year}
  {2012})}\BibitemShut {NoStop}%
\bibitem [{\citenamefont {Moore}\ and\ \citenamefont {Read}(1991)}]{fracjos1}%
  \BibitemOpen
  \bibfield  {author} {\bibinfo {author} {\bibfnamefont {G.}~\bibnamefont
  {Moore}}\ and\ \bibinfo {author} {\bibfnamefont {N.}~\bibnamefont {Read}},\
  }\bibfield  {title} {\bibinfo {title} {Nonabelions in the fractional quantum
  hall effect},\ }\href
  {https://doi.org/https://doi.org/10.1016/0550-3213(91)90407-O} {\bibfield
  {journal} {\bibinfo  {journal} {Nuclear Physics B}\ }\textbf {\bibinfo
  {volume} {360}},\ \bibinfo {pages} {362} (\bibinfo {year}
  {1991})}\BibitemShut {NoStop}%
\bibitem [{\citenamefont {Read}\ and\ \citenamefont {Green}(2000)}]{fracjos2}%
  \BibitemOpen
  \bibfield  {author} {\bibinfo {author} {\bibfnamefont {N.}~\bibnamefont
  {Read}}\ and\ \bibinfo {author} {\bibfnamefont {D.}~\bibnamefont {Green}},\
  }\bibfield  {title} {\bibinfo {title} {Paired states of fermions in two
  dimensions with breaking of parity and time-reversal symmetries and the
  fractional quantum hall effect},\ }\href
  {https://doi.org/10.1103/PhysRevB.61.10267} {\bibfield  {journal} {\bibinfo
  {journal} {Phys. Rev. B}\ }\textbf {\bibinfo {volume} {61}},\ \bibinfo
  {pages} {10267} (\bibinfo {year} {2000})}\BibitemShut {NoStop}%
\bibitem [{\citenamefont {Chiu}\ and\ \citenamefont
  {Das~Sarma}(2019)}]{fracjos3}%
  \BibitemOpen
  \bibfield  {author} {\bibinfo {author} {\bibfnamefont {C.-K.}\ \bibnamefont
  {Chiu}}\ and\ \bibinfo {author} {\bibfnamefont {S.}~\bibnamefont
  {Das~Sarma}},\ }\bibfield  {title} {\bibinfo {title} {Fractional josephson
  effect with and without majorana zero modes},\ }\href
  {https://doi.org/10.1103/PhysRevB.99.035312} {\bibfield  {journal} {\bibinfo
  {journal} {Phys. Rev. B}\ }\textbf {\bibinfo {volume} {99}},\ \bibinfo
  {pages} {035312} (\bibinfo {year} {2019})}\BibitemShut {NoStop}%
\bibitem [{\citenamefont {Kwon}\ \emph {et~al.}(2003)\citenamefont {Kwon},
  \citenamefont {Sengupta},\ and\ \citenamefont {Yakovenko}}]{Majref7}%
  \BibitemOpen
  \bibfield  {author} {\bibinfo {author} {\bibfnamefont {H.-J.}\ \bibnamefont
  {Kwon}}, \bibinfo {author} {\bibfnamefont {K.}~\bibnamefont {Sengupta}},\
  and\ \bibinfo {author} {\bibfnamefont {V.~M.}\ \bibnamefont {Yakovenko}},\
  }\bibfield  {title} {\bibinfo {title} {Fractional ac josephson effect in p-
  and d-wave superconductors},\ }\href
  {https://doi.org/10.1140/epjb/e2004-00066-4} {\bibfield  {journal} {\bibinfo
  {journal} {The European Physical Journal B - Condensed Matter}\ }\textbf
  {\bibinfo {volume} {37}},\ \bibinfo {pages} {349} (\bibinfo {year}
  {2003})}\BibitemShut {NoStop}%
\bibitem [{\citenamefont {Lutchyn}\ \emph {et~al.}(2010)\citenamefont
  {Lutchyn}, \citenamefont {Sau},\ and\ \citenamefont {Das~Sarma}}]{MajreF9}%
  \BibitemOpen
  \bibfield  {author} {\bibinfo {author} {\bibfnamefont {R.~M.}\ \bibnamefont
  {Lutchyn}}, \bibinfo {author} {\bibfnamefont {J.~D.}\ \bibnamefont {Sau}},\
  and\ \bibinfo {author} {\bibfnamefont {S.}~\bibnamefont {Das~Sarma}},\
  }\bibfield  {title} {\bibinfo {title} {Majorana fermions and a topological
  phase transition in semiconductor-superconductor heterostructures},\ }\href
  {https://doi.org/10.1103/PhysRevLett.105.077001} {\bibfield  {journal}
  {\bibinfo  {journal} {Phys. Rev. Lett.}\ }\textbf {\bibinfo {volume} {105}},\
  \bibinfo {pages} {077001} (\bibinfo {year} {2010})}\BibitemShut {NoStop}%
\bibitem [{\citenamefont {Oka}\ and\ \citenamefont
  {Aoki}(2009{\natexlab{a}})}]{Floref1}%
  \BibitemOpen
  \bibfield  {author} {\bibinfo {author} {\bibfnamefont {T.}~\bibnamefont
  {Oka}}\ and\ \bibinfo {author} {\bibfnamefont {H.}~\bibnamefont {Aoki}},\
  }\bibfield  {title} {\bibinfo {title} {Photovoltaic hall effect in
  graphene},\ }\href {https://doi.org/10.1103/PhysRevB.79.081406} {\bibfield
  {journal} {\bibinfo  {journal} {Phys. Rev. B}\ }\textbf {\bibinfo {volume}
  {79}},\ \bibinfo {pages} {081406} (\bibinfo {year}
  {2009}{\natexlab{a}})}\BibitemShut {NoStop}%
\bibitem [{\citenamefont {Perez-Piskunow}\ \emph
  {et~al.}(2014{\natexlab{a}})\citenamefont {Perez-Piskunow}, \citenamefont
  {Usaj}, \citenamefont {Balseiro},\ and\ \citenamefont {Torres}}]{Floref4}%
  \BibitemOpen
  \bibfield  {author} {\bibinfo {author} {\bibfnamefont {P.~M.}\ \bibnamefont
  {Perez-Piskunow}}, \bibinfo {author} {\bibfnamefont {G.}~\bibnamefont
  {Usaj}}, \bibinfo {author} {\bibfnamefont {C.~A.}\ \bibnamefont {Balseiro}},\
  and\ \bibinfo {author} {\bibfnamefont {L.~E. F.~F.}\ \bibnamefont {Torres}},\
  }\bibfield  {title} {\bibinfo {title} {Floquet chiral edge states in
  graphene},\ }\href {https://doi.org/10.1103/PhysRevB.89.121401} {\bibfield
  {journal} {\bibinfo  {journal} {Phys. Rev. B}\ }\textbf {\bibinfo {volume}
  {89}},\ \bibinfo {pages} {121401} (\bibinfo {year}
  {2014}{\natexlab{a}})}\BibitemShut {NoStop}%
\bibitem [{\citenamefont {Fidkowski}\ \emph {et~al.}(2019)\citenamefont
  {Fidkowski}, \citenamefont {Po}, \citenamefont {Potter},\ and\ \citenamefont
  {Vishwanath}}]{Floref5}%
  \BibitemOpen
  \bibfield  {author} {\bibinfo {author} {\bibfnamefont {L.}~\bibnamefont
  {Fidkowski}}, \bibinfo {author} {\bibfnamefont {H.~C.}\ \bibnamefont {Po}},
  \bibinfo {author} {\bibfnamefont {A.~C.}\ \bibnamefont {Potter}},\ and\
  \bibinfo {author} {\bibfnamefont {A.}~\bibnamefont {Vishwanath}},\ }\bibfield
   {title} {\bibinfo {title} {Interacting invariants for floquet phases of
  fermions in two dimensions},\ }\href
  {https://doi.org/10.1103/PhysRevB.99.085115} {\bibfield  {journal} {\bibinfo
  {journal} {Phys. Rev. B}\ }\textbf {\bibinfo {volume} {99}},\ \bibinfo
  {pages} {085115} (\bibinfo {year} {2019})}\BibitemShut {NoStop}%
\bibitem [{\citenamefont {Harper}\ \emph {et~al.}(2020)\citenamefont {Harper},
  \citenamefont {Roy}, \citenamefont {Rudner},\ and\ \citenamefont
  {Sondhi}}]{Floref6}%
  \BibitemOpen
  \bibfield  {author} {\bibinfo {author} {\bibfnamefont {F.}~\bibnamefont
  {Harper}}, \bibinfo {author} {\bibfnamefont {R.}~\bibnamefont {Roy}},
  \bibinfo {author} {\bibfnamefont {M.~S.}\ \bibnamefont {Rudner}},\ and\
  \bibinfo {author} {\bibfnamefont {S.}~\bibnamefont {Sondhi}},\ }\bibfield
  {title} {\bibinfo {title} {Topology and broken symmetry in floquet systems},\
  }\href {https://doi.org/10.1146/annurev-conmatphys-031218-013721} {\bibfield
  {journal} {\bibinfo  {journal} {Annual Review of Condensed Matter Physics}\
  }\textbf {\bibinfo {volume} {11}},\ \bibinfo {pages} {345} (\bibinfo {year}
  {2020})}\BibitemShut {NoStop}%
\bibitem [{\citenamefont {Zhang}\ and\ \citenamefont {Yang}(2021)}]{Floref7}%
  \BibitemOpen
  \bibfield  {author} {\bibinfo {author} {\bibfnamefont {R.-X.}\ \bibnamefont
  {Zhang}}\ and\ \bibinfo {author} {\bibfnamefont {Z.-C.}\ \bibnamefont
  {Yang}},\ }\bibfield  {title} {\bibinfo {title} {Tunable fragile topology in
  floquet systems},\ }\bibfield  {journal} {\bibinfo  {journal} {Physical
  Review B}\ }\textbf {\bibinfo {volume} {103}},\ \href
  {https://doi.org/10.1103/physrevb.103.l121115} {10.1103/physrevb.103.l121115}
  (\bibinfo {year} {2021})\BibitemShut {NoStop}%
\bibitem [{\citenamefont {Kitagawa}\ \emph
  {et~al.}(2010{\natexlab{a}})\citenamefont {Kitagawa}, \citenamefont {Berg},
  \citenamefont {Rudner},\ and\ \citenamefont {Demler}}]{Floref8}%
  \BibitemOpen
  \bibfield  {author} {\bibinfo {author} {\bibfnamefont {T.}~\bibnamefont
  {Kitagawa}}, \bibinfo {author} {\bibfnamefont {E.}~\bibnamefont {Berg}},
  \bibinfo {author} {\bibfnamefont {M.}~\bibnamefont {Rudner}},\ and\ \bibinfo
  {author} {\bibfnamefont {E.}~\bibnamefont {Demler}},\ }\bibfield  {title}
  {\bibinfo {title} {Topological characterization of periodically driven
  quantum systems},\ }\href {https://doi.org/10.1103/PhysRevB.82.235114}
  {\bibfield  {journal} {\bibinfo  {journal} {Phys. Rev. B}\ }\textbf {\bibinfo
  {volume} {82}},\ \bibinfo {pages} {235114} (\bibinfo {year}
  {2010}{\natexlab{a}})}\BibitemShut {NoStop}%
\bibitem [{\citenamefont {Rudner}\ \emph {et~al.}(2013)\citenamefont {Rudner},
  \citenamefont {Lindner}, \citenamefont {Berg},\ and\ \citenamefont
  {Levin}}]{Floref9}%
  \BibitemOpen
  \bibfield  {author} {\bibinfo {author} {\bibfnamefont {M.~S.}\ \bibnamefont
  {Rudner}}, \bibinfo {author} {\bibfnamefont {N.~H.}\ \bibnamefont {Lindner}},
  \bibinfo {author} {\bibfnamefont {E.}~\bibnamefont {Berg}},\ and\ \bibinfo
  {author} {\bibfnamefont {M.}~\bibnamefont {Levin}},\ }\bibfield  {title}
  {\bibinfo {title} {Anomalous edge states and the bulk-edge correspondence for
  periodically driven two-dimensional systems},\ }\href
  {https://doi.org/10.1103/PhysRevX.3.031005} {\bibfield  {journal} {\bibinfo
  {journal} {Phys. Rev. X}\ }\textbf {\bibinfo {volume} {3}},\ \bibinfo {pages}
  {031005} (\bibinfo {year} {2013})}\BibitemShut {NoStop}%
\bibitem [{\citenamefont {Nathan}\ and\ \citenamefont
  {Rudner}(2015)}]{Floref10}%
  \BibitemOpen
  \bibfield  {author} {\bibinfo {author} {\bibfnamefont {F.}~\bibnamefont
  {Nathan}}\ and\ \bibinfo {author} {\bibfnamefont {M.~S.}\ \bibnamefont
  {Rudner}},\ }\bibfield  {title} {\bibinfo {title} {Topological singularities
  and the general classification of floquet–bloch systems},\ }\href
  {https://doi.org/10.1088/1367-2630/17/12/125014} {\bibfield  {journal}
  {\bibinfo  {journal} {New Journal of Physics}\ }\textbf {\bibinfo {volume}
  {17}},\ \bibinfo {pages} {125014} (\bibinfo {year} {2015})}\BibitemShut
  {NoStop}%
\bibitem [{\citenamefont {Perez-Piskunow}\ \emph {et~al.}(2015)\citenamefont
  {Perez-Piskunow}, \citenamefont {Foa~Torres},\ and\ \citenamefont
  {Usaj}}]{Floref11}%
  \BibitemOpen
  \bibfield  {author} {\bibinfo {author} {\bibfnamefont {P.~M.}\ \bibnamefont
  {Perez-Piskunow}}, \bibinfo {author} {\bibfnamefont {L.~E.~F.}\ \bibnamefont
  {Foa~Torres}},\ and\ \bibinfo {author} {\bibfnamefont {G.}~\bibnamefont
  {Usaj}},\ }\bibfield  {title} {\bibinfo {title} {Hierarchy of floquet gaps
  and edge states for driven honeycomb lattices},\ }\href
  {https://doi.org/10.1103/PhysRevA.91.043625} {\bibfield  {journal} {\bibinfo
  {journal} {Phys. Rev. A}\ }\textbf {\bibinfo {volume} {91}},\ \bibinfo
  {pages} {043625} (\bibinfo {year} {2015})}\BibitemShut {NoStop}%
\bibitem [{\citenamefont {Morimoto}\ \emph {et~al.}(2017)\citenamefont
  {Morimoto}, \citenamefont {Po},\ and\ \citenamefont {Vishwanath}}]{Floref12}%
  \BibitemOpen
  \bibfield  {author} {\bibinfo {author} {\bibfnamefont {T.}~\bibnamefont
  {Morimoto}}, \bibinfo {author} {\bibfnamefont {H.~C.}\ \bibnamefont {Po}},\
  and\ \bibinfo {author} {\bibfnamefont {A.}~\bibnamefont {Vishwanath}},\
  }\bibfield  {title} {\bibinfo {title} {Floquet topological phases protected
  by time glide symmetry},\ }\href {https://doi.org/10.1103/PhysRevB.95.195155}
  {\bibfield  {journal} {\bibinfo  {journal} {Phys. Rev. B}\ }\textbf {\bibinfo
  {volume} {95}},\ \bibinfo {pages} {195155} (\bibinfo {year}
  {2017})}\BibitemShut {NoStop}%
\bibitem [{\citenamefont {Molignini}\ \emph {et~al.}(2020)\citenamefont
  {Molignini}, \citenamefont {Chen},\ and\ \citenamefont {Chitra}}]{Floref13}%
  \BibitemOpen
  \bibfield  {author} {\bibinfo {author} {\bibfnamefont {P.}~\bibnamefont
  {Molignini}}, \bibinfo {author} {\bibfnamefont {W.}~\bibnamefont {Chen}},\
  and\ \bibinfo {author} {\bibfnamefont {R.}~\bibnamefont {Chitra}},\
  }\bibfield  {title} {\bibinfo {title} {Generating quantum multicriticality in
  topological insulators by periodic driving},\ }\href
  {https://doi.org/10.1103/PhysRevB.101.165106} {\bibfield  {journal} {\bibinfo
   {journal} {Phys. Rev. B}\ }\textbf {\bibinfo {volume} {101}},\ \bibinfo
  {pages} {165106} (\bibinfo {year} {2020})}\BibitemShut {NoStop}%
\bibitem [{\citenamefont {Zhang}\ and\ \citenamefont {Yang}(2020)}]{Floref14}%
  \BibitemOpen
  \bibfield  {author} {\bibinfo {author} {\bibfnamefont {R.-X.}\ \bibnamefont
  {Zhang}}\ and\ \bibinfo {author} {\bibfnamefont {Z.-C.}\ \bibnamefont
  {Yang}},\ }\href {https://doi.org/10.48550/ARXIV.2010.07945} {\bibinfo
  {title} {Theory of anomalous floquet higher-order topology: Classification,
  characterization, and bulk-boundary correspondence}} (\bibinfo {year}
  {2020})\BibitemShut {NoStop}%
\bibitem [{\citenamefont {G\'omez-Le\'on}\ and\ \citenamefont
  {Platero}(2013)}]{Floref15}%
  \BibitemOpen
  \bibfield  {author} {\bibinfo {author} {\bibfnamefont {A.}~\bibnamefont
  {G\'omez-Le\'on}}\ and\ \bibinfo {author} {\bibfnamefont {G.}~\bibnamefont
  {Platero}},\ }\bibfield  {title} {\bibinfo {title} {Floquet-bloch theory and
  topology in periodically driven lattices},\ }\href
  {https://doi.org/10.1103/PhysRevLett.110.200403} {\bibfield  {journal}
  {\bibinfo  {journal} {Phys. Rev. Lett.}\ }\textbf {\bibinfo {volume} {110}},\
  \bibinfo {pages} {200403} (\bibinfo {year} {2013})}\BibitemShut {NoStop}%
\bibitem [{\citenamefont {Lababidi}\ \emph {et~al.}(2014)\citenamefont
  {Lababidi}, \citenamefont {Satija},\ and\ \citenamefont {Zhao}}]{Floref16}%
  \BibitemOpen
  \bibfield  {author} {\bibinfo {author} {\bibfnamefont {M.}~\bibnamefont
  {Lababidi}}, \bibinfo {author} {\bibfnamefont {I.~I.}\ \bibnamefont
  {Satija}},\ and\ \bibinfo {author} {\bibfnamefont {E.}~\bibnamefont {Zhao}},\
  }\bibfield  {title} {\bibinfo {title} {Counter-propagating edge modes and
  topological phases of a kicked quantum hall system},\ }\href
  {https://doi.org/10.1103/PhysRevLett.112.026805} {\bibfield  {journal}
  {\bibinfo  {journal} {Phys. Rev. Lett.}\ }\textbf {\bibinfo {volume} {112}},\
  \bibinfo {pages} {026805} (\bibinfo {year} {2014})}\BibitemShut {NoStop}%
\bibitem [{\citenamefont {Perez-Piskunow}\ \emph
  {et~al.}(2014{\natexlab{b}})\citenamefont {Perez-Piskunow}, \citenamefont
  {Usaj}, \citenamefont {Balseiro},\ and\ \citenamefont {Torres}}]{Floref17}%
  \BibitemOpen
  \bibfield  {author} {\bibinfo {author} {\bibfnamefont {P.~M.}\ \bibnamefont
  {Perez-Piskunow}}, \bibinfo {author} {\bibfnamefont {G.}~\bibnamefont
  {Usaj}}, \bibinfo {author} {\bibfnamefont {C.~A.}\ \bibnamefont {Balseiro}},\
  and\ \bibinfo {author} {\bibfnamefont {L.~E. F.~F.}\ \bibnamefont {Torres}},\
  }\bibfield  {title} {\bibinfo {title} {Floquet chiral edge states in
  graphene},\ }\href {https://doi.org/10.1103/PhysRevB.89.121401} {\bibfield
  {journal} {\bibinfo  {journal} {Phys. Rev. B}\ }\textbf {\bibinfo {volume}
  {89}},\ \bibinfo {pages} {121401} (\bibinfo {year}
  {2014}{\natexlab{b}})}\BibitemShut {NoStop}%
\bibitem [{\citenamefont {Eckardt}(2017)}]{Floref18}%
  \BibitemOpen
  \bibfield  {author} {\bibinfo {author} {\bibfnamefont {A.}~\bibnamefont
  {Eckardt}},\ }\bibfield  {title} {\bibinfo {title} {Colloquium: Atomic
  quantum gases in periodically driven optical lattices},\ }\href
  {https://doi.org/10.1103/RevModPhys.89.011004} {\bibfield  {journal}
  {\bibinfo  {journal} {Rev. Mod. Phys.}\ }\textbf {\bibinfo {volume} {89}},\
  \bibinfo {pages} {011004} (\bibinfo {year} {2017})}\BibitemShut {NoStop}%
\bibitem [{\citenamefont {Wang}\ \emph {et~al.}(2013)\citenamefont {Wang},
  \citenamefont {Steinberg}, \citenamefont {Jarillo-Herrero},\ and\
  \citenamefont {Gedik}}]{flo_solid1}%
  \BibitemOpen
  \bibfield  {author} {\bibinfo {author} {\bibfnamefont {Y.~H.}\ \bibnamefont
  {Wang}}, \bibinfo {author} {\bibfnamefont {H.}~\bibnamefont {Steinberg}},
  \bibinfo {author} {\bibfnamefont {P.}~\bibnamefont {Jarillo-Herrero}},\ and\
  \bibinfo {author} {\bibfnamefont {N.}~\bibnamefont {Gedik}},\ }\bibfield
  {title} {\bibinfo {title} {Observation of floquet-bloch states on the surface
  of a topological insulator},\ }\href
  {https://doi.org/10.1126/science.1239834} {\bibfield  {journal} {\bibinfo
  {journal} {Science}\ }\textbf {\bibinfo {volume} {342}},\ \bibinfo {pages}
  {453} (\bibinfo {year} {2013})}\BibitemShut {NoStop}%
\bibitem [{\citenamefont {McIver}\ \emph {et~al.}(2019)\citenamefont {McIver},
  \citenamefont {Schulte}, \citenamefont {Stein}, \citenamefont {Matsuyama},
  \citenamefont {Jotzu}, \citenamefont {Meier},\ and\ \citenamefont
  {Cavalleri}}]{flo_solid2}%
  \BibitemOpen
  \bibfield  {author} {\bibinfo {author} {\bibfnamefont {J.~W.}\ \bibnamefont
  {McIver}}, \bibinfo {author} {\bibfnamefont {B.}~\bibnamefont {Schulte}},
  \bibinfo {author} {\bibfnamefont {F.-U.}\ \bibnamefont {Stein}}, \bibinfo
  {author} {\bibfnamefont {T.}~\bibnamefont {Matsuyama}}, \bibinfo {author}
  {\bibfnamefont {G.}~\bibnamefont {Jotzu}}, \bibinfo {author} {\bibfnamefont
  {G.}~\bibnamefont {Meier}},\ and\ \bibinfo {author} {\bibfnamefont
  {A.}~\bibnamefont {Cavalleri}},\ }\bibfield  {title} {\bibinfo {title}
  {Light-induced anomalous hall effect in graphene},\ }\href
  {https://doi.org/10.1038/s41567-019-0698-y} {\bibfield  {journal} {\bibinfo
  {journal} {Nature Physics}\ }\textbf {\bibinfo {volume} {16}},\ \bibinfo
  {pages} {38} (\bibinfo {year} {2019})}\BibitemShut {NoStop}%
\bibitem [{\citenamefont {Jotzu}\ \emph {et~al.}(2014)\citenamefont {Jotzu},
  \citenamefont {Messer}, \citenamefont {Desbuquois}, \citenamefont {Lebrat},
  \citenamefont {Uehlinger}, \citenamefont {Greif},\ and\ \citenamefont
  {Esslinger}}]{flo_cold1}%
  \BibitemOpen
  \bibfield  {author} {\bibinfo {author} {\bibfnamefont {G.}~\bibnamefont
  {Jotzu}}, \bibinfo {author} {\bibfnamefont {M.}~\bibnamefont {Messer}},
  \bibinfo {author} {\bibfnamefont {R.}~\bibnamefont {Desbuquois}}, \bibinfo
  {author} {\bibfnamefont {M.}~\bibnamefont {Lebrat}}, \bibinfo {author}
  {\bibfnamefont {T.}~\bibnamefont {Uehlinger}}, \bibinfo {author}
  {\bibfnamefont {D.}~\bibnamefont {Greif}},\ and\ \bibinfo {author}
  {\bibfnamefont {T.}~\bibnamefont {Esslinger}},\ }\bibfield  {title} {\bibinfo
  {title} {Experimental realization of the topological haldane model with
  ultracold fermions},\ }\href {https://doi.org/10.1038/nature13915} {\bibfield
   {journal} {\bibinfo  {journal} {Nature}\ }\textbf {\bibinfo {volume}
  {515}},\ \bibinfo {pages} {237} (\bibinfo {year} {2014})}\BibitemShut
  {NoStop}%
\bibitem [{\citenamefont {Fläschner}\ \emph {et~al.}(2016)\citenamefont
  {Fläschner}, \citenamefont {Rem}, \citenamefont {Tarnowski}, \citenamefont
  {Vogel}, \citenamefont {Lühmann}, \citenamefont {Sengstock},\ and\
  \citenamefont {Weitenberg}}]{flo_cold2}%
  \BibitemOpen
  \bibfield  {author} {\bibinfo {author} {\bibfnamefont {N.}~\bibnamefont
  {Fläschner}}, \bibinfo {author} {\bibfnamefont {B.~S.}\ \bibnamefont {Rem}},
  \bibinfo {author} {\bibfnamefont {M.}~\bibnamefont {Tarnowski}}, \bibinfo
  {author} {\bibfnamefont {D.}~\bibnamefont {Vogel}}, \bibinfo {author}
  {\bibfnamefont {D.-S.}\ \bibnamefont {Lühmann}}, \bibinfo {author}
  {\bibfnamefont {K.}~\bibnamefont {Sengstock}},\ and\ \bibinfo {author}
  {\bibfnamefont {C.}~\bibnamefont {Weitenberg}},\ }\bibfield  {title}
  {\bibinfo {title} {Experimental reconstruction of the berry curvature in a
  floquet bloch band},\ }\href {https://doi.org/10.1126/science.aad4568}
  {\bibfield  {journal} {\bibinfo  {journal} {Science}\ }\textbf {\bibinfo
  {volume} {352}},\ \bibinfo {pages} {1091} (\bibinfo {year} {2016})},\ \Eprint
  {https://arxiv.org/abs/https://www.science.org/doi/pdf/10.1126/science.aad4568}
  {https://www.science.org/doi/pdf/10.1126/science.aad4568} \BibitemShut
  {NoStop}%
\bibitem [{\citenamefont {Wintersperger}\ \emph {et~al.}(2020)\citenamefont
  {Wintersperger}, \citenamefont {Braun}, \citenamefont {Ünal}, \citenamefont
  {Eckardt}, \citenamefont {Liberto}, \citenamefont {Goldman}, \citenamefont
  {Bloch},\ and\ \citenamefont {Aidelsburger}}]{flo_cold3}%
  \BibitemOpen
  \bibfield  {author} {\bibinfo {author} {\bibfnamefont {K.}~\bibnamefont
  {Wintersperger}}, \bibinfo {author} {\bibfnamefont {C.}~\bibnamefont
  {Braun}}, \bibinfo {author} {\bibfnamefont {F.~N.}\ \bibnamefont {Ünal}},
  \bibinfo {author} {\bibfnamefont {A.}~\bibnamefont {Eckardt}}, \bibinfo
  {author} {\bibfnamefont {M.~D.}\ \bibnamefont {Liberto}}, \bibinfo {author}
  {\bibfnamefont {N.}~\bibnamefont {Goldman}}, \bibinfo {author} {\bibfnamefont
  {I.}~\bibnamefont {Bloch}},\ and\ \bibinfo {author} {\bibfnamefont
  {M.}~\bibnamefont {Aidelsburger}},\ }\bibfield  {title} {\bibinfo {title}
  {Realization of an anomalous floquet topological system with ultracold
  atoms},\ }\href {https://doi.org/10.1038/s41567-020-0949-y} {\bibfield
  {journal} {\bibinfo  {journal} {Nature Physics}\ }\textbf {\bibinfo {volume}
  {16}},\ \bibinfo {pages} {1058} (\bibinfo {year} {2020})}\BibitemShut
  {NoStop}%
\bibitem [{\citenamefont {Liu}\ \emph {et~al.}(2019{\natexlab{a}})\citenamefont
  {Liu}, \citenamefont {Xiong}, \citenamefont {Zhang},\ and\ \citenamefont
  {An}}]{flo_cold4}%
  \BibitemOpen
  \bibfield  {author} {\bibinfo {author} {\bibfnamefont {H.}~\bibnamefont
  {Liu}}, \bibinfo {author} {\bibfnamefont {T.-S.}\ \bibnamefont {Xiong}},
  \bibinfo {author} {\bibfnamefont {W.}~\bibnamefont {Zhang}},\ and\ \bibinfo
  {author} {\bibfnamefont {J.-H.}\ \bibnamefont {An}},\ }\bibfield  {title}
  {\bibinfo {title} {Floquet engineering of exotic topological phases in
  systems of cold atoms},\ }\bibfield  {journal} {\bibinfo  {journal} {Physical
  Review A}\ }\textbf {\bibinfo {volume} {100}},\ \href
  {https://doi.org/10.1103/physreva.100.023622} {10.1103/physreva.100.023622}
  (\bibinfo {year} {2019}{\natexlab{a}})\BibitemShut {NoStop}%
\bibitem [{\citenamefont {Potirniche}\ \emph {et~al.}(2017)\citenamefont
  {Potirniche}, \citenamefont {Potter}, \citenamefont {Schleier-Smith},
  \citenamefont {Vishwanath},\ and\ \citenamefont {Yao}}]{flo_cold5}%
  \BibitemOpen
  \bibfield  {author} {\bibinfo {author} {\bibfnamefont {I.-D.}\ \bibnamefont
  {Potirniche}}, \bibinfo {author} {\bibfnamefont {A.~C.}\ \bibnamefont
  {Potter}}, \bibinfo {author} {\bibfnamefont {M.}~\bibnamefont
  {Schleier-Smith}}, \bibinfo {author} {\bibfnamefont {A.}~\bibnamefont
  {Vishwanath}},\ and\ \bibinfo {author} {\bibfnamefont {N.~Y.}\ \bibnamefont
  {Yao}},\ }\bibfield  {title} {\bibinfo {title} {Floquet symmetry-protected
  topological phases in cold-atom systems},\ }\href
  {https://doi.org/10.1103/PhysRevLett.119.123601} {\bibfield  {journal}
  {\bibinfo  {journal} {Phys. Rev. Lett.}\ }\textbf {\bibinfo {volume} {119}},\
  \bibinfo {pages} {123601} (\bibinfo {year} {2017})}\BibitemShut {NoStop}%
\bibitem [{\citenamefont {Rechtsman}\ \emph
  {et~al.}(2013{\natexlab{a}})\citenamefont {Rechtsman}, \citenamefont
  {Zeuner}, \citenamefont {Plotnik}, \citenamefont {Lumer}, \citenamefont
  {Podolsky}, \citenamefont {Dreisow}, \citenamefont {Nolte}, \citenamefont
  {Segev},\ and\ \citenamefont {Szameit}}]{flo_opt1}%
  \BibitemOpen
  \bibfield  {author} {\bibinfo {author} {\bibfnamefont {M.~C.}\ \bibnamefont
  {Rechtsman}}, \bibinfo {author} {\bibfnamefont {J.~M.}\ \bibnamefont
  {Zeuner}}, \bibinfo {author} {\bibfnamefont {Y.}~\bibnamefont {Plotnik}},
  \bibinfo {author} {\bibfnamefont {Y.}~\bibnamefont {Lumer}}, \bibinfo
  {author} {\bibfnamefont {D.}~\bibnamefont {Podolsky}}, \bibinfo {author}
  {\bibfnamefont {F.}~\bibnamefont {Dreisow}}, \bibinfo {author} {\bibfnamefont
  {S.}~\bibnamefont {Nolte}}, \bibinfo {author} {\bibfnamefont
  {M.}~\bibnamefont {Segev}},\ and\ \bibinfo {author} {\bibfnamefont
  {A.}~\bibnamefont {Szameit}},\ }\bibfield  {title} {\bibinfo {title}
  {Photonic floquet topological insulators},\ }\href
  {https://doi.org/10.1038/nature12066} {\bibfield  {journal} {\bibinfo
  {journal} {Nature}\ }\textbf {\bibinfo {volume} {496}},\ \bibinfo {pages}
  {196} (\bibinfo {year} {2013}{\natexlab{a}})}\BibitemShut {NoStop}%
\bibitem [{\citenamefont {Afzal}\ \emph
  {et~al.}(2020{\natexlab{a}})\citenamefont {Afzal}, \citenamefont
  {Zimmerling}, \citenamefont {Ren}, \citenamefont {Perron},\ and\
  \citenamefont {Van}}]{flo_opt2}%
  \BibitemOpen
  \bibfield  {author} {\bibinfo {author} {\bibfnamefont {S.}~\bibnamefont
  {Afzal}}, \bibinfo {author} {\bibfnamefont {T.~J.}\ \bibnamefont
  {Zimmerling}}, \bibinfo {author} {\bibfnamefont {Y.}~\bibnamefont {Ren}},
  \bibinfo {author} {\bibfnamefont {D.}~\bibnamefont {Perron}},\ and\ \bibinfo
  {author} {\bibfnamefont {V.}~\bibnamefont {Van}},\ }\bibfield  {title}
  {\bibinfo {title} {Realization of anomalous floquet insulators in strongly
  coupled nanophotonic lattices},\ }\href
  {https://doi.org/10.1103/PhysRevLett.124.253601} {\bibfield  {journal}
  {\bibinfo  {journal} {Phys. Rev. Lett.}\ }\textbf {\bibinfo {volume} {124}},\
  \bibinfo {pages} {253601} (\bibinfo {year} {2020}{\natexlab{a}})}\BibitemShut
  {NoStop}%
\bibitem [{\citenamefont {Mukherjee}\ and\ \citenamefont
  {Rechtsman}(2020)}]{flo_opt3}%
  \BibitemOpen
  \bibfield  {author} {\bibinfo {author} {\bibfnamefont {S.}~\bibnamefont
  {Mukherjee}}\ and\ \bibinfo {author} {\bibfnamefont {M.~C.}\ \bibnamefont
  {Rechtsman}},\ }\bibfield  {title} {\bibinfo {title} {Observation of floquet
  solitons in a topological bandgap},\ }\href
  {https://doi.org/10.1126/science.aba8725} {\bibfield  {journal} {\bibinfo
  {journal} {Science}\ }\textbf {\bibinfo {volume} {368}},\ \bibinfo {pages}
  {856} (\bibinfo {year} {2020})},\ \Eprint
  {https://arxiv.org/abs/https://www.science.org/doi/pdf/10.1126/science.aba8725}
  {https://www.science.org/doi/pdf/10.1126/science.aba8725} \BibitemShut
  {NoStop}%
\bibitem [{\citenamefont {Kundu}\ and\ \citenamefont
  {Seradjeh}(2013)}]{zeropeak2a}%
  \BibitemOpen
  \bibfield  {author} {\bibinfo {author} {\bibfnamefont {A.}~\bibnamefont
  {Kundu}}\ and\ \bibinfo {author} {\bibfnamefont {B.}~\bibnamefont
  {Seradjeh}},\ }\bibfield  {title} {\bibinfo {title} {Transport signatures of
  floquet majorana fermions in driven topological superconductors},\ }\href
  {https://doi.org/10.1103/PhysRevLett.111.136402} {\bibfield  {journal}
  {\bibinfo  {journal} {Phys. Rev. Lett.}\ }\textbf {\bibinfo {volume} {111}},\
  \bibinfo {pages} {136402} (\bibinfo {year} {2013})}\BibitemShut {NoStop}%
\bibitem [{\citenamefont {Yao}\ \emph {et~al.}(2017)\citenamefont {Yao},
  \citenamefont {Yan},\ and\ \citenamefont {Wang}}]{floq_maj1}%
  \BibitemOpen
  \bibfield  {author} {\bibinfo {author} {\bibfnamefont {S.}~\bibnamefont
  {Yao}}, \bibinfo {author} {\bibfnamefont {Z.}~\bibnamefont {Yan}},\ and\
  \bibinfo {author} {\bibfnamefont {Z.}~\bibnamefont {Wang}},\ }\bibfield
  {title} {\bibinfo {title} {Topological invariants of floquet systems: General
  formulation, special properties, and floquet topological defects},\ }\href
  {https://doi.org/10.1103/PhysRevB.96.195303} {\bibfield  {journal} {\bibinfo
  {journal} {Phys. Rev. B}\ }\textbf {\bibinfo {volume} {96}},\ \bibinfo
  {pages} {195303} (\bibinfo {year} {2017})}\BibitemShut {NoStop}%
\bibitem [{\citenamefont {Jiang}\ \emph {et~al.}(2011)\citenamefont {Jiang},
  \citenamefont {Kitagawa}, \citenamefont {Alicea}, \citenamefont {Akhmerov},
  \citenamefont {Pekker}, \citenamefont {Refael}, \citenamefont {Cirac},
  \citenamefont {Demler}, \citenamefont {Lukin},\ and\ \citenamefont
  {Zoller}}]{floq_maj2}%
  \BibitemOpen
  \bibfield  {author} {\bibinfo {author} {\bibfnamefont {L.}~\bibnamefont
  {Jiang}}, \bibinfo {author} {\bibfnamefont {T.}~\bibnamefont {Kitagawa}},
  \bibinfo {author} {\bibfnamefont {J.}~\bibnamefont {Alicea}}, \bibinfo
  {author} {\bibfnamefont {A.~R.}\ \bibnamefont {Akhmerov}}, \bibinfo {author}
  {\bibfnamefont {D.}~\bibnamefont {Pekker}}, \bibinfo {author} {\bibfnamefont
  {G.}~\bibnamefont {Refael}}, \bibinfo {author} {\bibfnamefont {J.~I.}\
  \bibnamefont {Cirac}}, \bibinfo {author} {\bibfnamefont {E.}~\bibnamefont
  {Demler}}, \bibinfo {author} {\bibfnamefont {M.~D.}\ \bibnamefont {Lukin}},\
  and\ \bibinfo {author} {\bibfnamefont {P.}~\bibnamefont {Zoller}},\
  }\bibfield  {title} {\bibinfo {title} {Majorana fermions in equilibrium and
  in driven cold-atom quantum wires},\ }\href
  {https://doi.org/10.1103/PhysRevLett.106.220402} {\bibfield  {journal}
  {\bibinfo  {journal} {Phys. Rev. Lett.}\ }\textbf {\bibinfo {volume} {106}},\
  \bibinfo {pages} {220402} (\bibinfo {year} {2011})}\BibitemShut {NoStop}%
\bibitem [{\citenamefont {Oka}\ and\ \citenamefont
  {Aoki}(2009{\natexlab{b}})}]{Floquetref1}%
  \BibitemOpen
  \bibfield  {author} {\bibinfo {author} {\bibfnamefont {T.}~\bibnamefont
  {Oka}}\ and\ \bibinfo {author} {\bibfnamefont {H.}~\bibnamefont {Aoki}},\
  }\bibfield  {title} {\bibinfo {title} {Photovoltaic hall effect in
  graphene},\ }\href {https://doi.org/10.1103/PhysRevB.79.081406} {\bibfield
  {journal} {\bibinfo  {journal} {Phys. Rev. B}\ }\textbf {\bibinfo {volume}
  {79}},\ \bibinfo {pages} {081406} (\bibinfo {year}
  {2009}{\natexlab{b}})}\BibitemShut {NoStop}%
\bibitem [{\citenamefont {Kitagawa}\ \emph
  {et~al.}(2010{\natexlab{b}})\citenamefont {Kitagawa}, \citenamefont {Berg},
  \citenamefont {Rudner},\ and\ \citenamefont {Demler}}]{Floquetref2}%
  \BibitemOpen
  \bibfield  {author} {\bibinfo {author} {\bibfnamefont {T.}~\bibnamefont
  {Kitagawa}}, \bibinfo {author} {\bibfnamefont {E.}~\bibnamefont {Berg}},
  \bibinfo {author} {\bibfnamefont {M.}~\bibnamefont {Rudner}},\ and\ \bibinfo
  {author} {\bibfnamefont {E.}~\bibnamefont {Demler}},\ }\bibfield  {title}
  {\bibinfo {title} {Topological characterization of periodically driven
  quantum systems},\ }\href {https://doi.org/10.1103/PhysRevB.82.235114}
  {\bibfield  {journal} {\bibinfo  {journal} {Phys. Rev. B}\ }\textbf {\bibinfo
  {volume} {82}},\ \bibinfo {pages} {235114} (\bibinfo {year}
  {2010}{\natexlab{b}})}\BibitemShut {NoStop}%
\bibitem [{\citenamefont {Gu}\ \emph {et~al.}(2011)\citenamefont {Gu},
  \citenamefont {Fertig}, \citenamefont {Arovas},\ and\ \citenamefont
  {Auerbach}}]{Floquetref3}%
  \BibitemOpen
  \bibfield  {author} {\bibinfo {author} {\bibfnamefont {Z.}~\bibnamefont
  {Gu}}, \bibinfo {author} {\bibfnamefont {H.~A.}\ \bibnamefont {Fertig}},
  \bibinfo {author} {\bibfnamefont {D.~P.}\ \bibnamefont {Arovas}},\ and\
  \bibinfo {author} {\bibfnamefont {A.}~\bibnamefont {Auerbach}},\ }\bibfield
  {title} {\bibinfo {title} {Floquet spectrum and transport through an
  irradiated graphene ribbon},\ }\href
  {https://doi.org/10.1103/PhysRevLett.107.216601} {\bibfield  {journal}
  {\bibinfo  {journal} {Phys. Rev. Lett.}\ }\textbf {\bibinfo {volume} {107}},\
  \bibinfo {pages} {216601} (\bibinfo {year} {2011})}\BibitemShut {NoStop}%
\bibitem [{\citenamefont {Kundu}\ \emph {et~al.}(2020)\citenamefont {Kundu},
  \citenamefont {Rudner}, \citenamefont {Berg},\ and\ \citenamefont
  {Lindner}}]{Floquetref4}%
  \BibitemOpen
  \bibfield  {author} {\bibinfo {author} {\bibfnamefont {A.}~\bibnamefont
  {Kundu}}, \bibinfo {author} {\bibfnamefont {M.}~\bibnamefont {Rudner}},
  \bibinfo {author} {\bibfnamefont {E.}~\bibnamefont {Berg}},\ and\ \bibinfo
  {author} {\bibfnamefont {N.~H.}\ \bibnamefont {Lindner}},\ }\bibfield
  {title} {\bibinfo {title} {Quantized large-bias current in the anomalous
  floquet-anderson insulator},\ }\href
  {https://doi.org/10.1103/PhysRevB.101.041403} {\bibfield  {journal} {\bibinfo
   {journal} {Phys. Rev. B}\ }\textbf {\bibinfo {volume} {101}},\ \bibinfo
  {pages} {041403} (\bibinfo {year} {2020})}\BibitemShut {NoStop}%
\bibitem [{\citenamefont {Kundu}\ \emph {et~al.}(2016)\citenamefont {Kundu},
  \citenamefont {Fertig},\ and\ \citenamefont {Seradjeh}}]{Floquetref5}%
  \BibitemOpen
  \bibfield  {author} {\bibinfo {author} {\bibfnamefont {A.}~\bibnamefont
  {Kundu}}, \bibinfo {author} {\bibfnamefont {H.~A.}\ \bibnamefont {Fertig}},\
  and\ \bibinfo {author} {\bibfnamefont {B.}~\bibnamefont {Seradjeh}},\
  }\bibfield  {title} {\bibinfo {title} {Floquet-engineered valleytronics in
  dirac systems},\ }\href {https://doi.org/10.1103/PhysRevLett.116.016802}
  {\bibfield  {journal} {\bibinfo  {journal} {Phys. Rev. Lett.}\ }\textbf
  {\bibinfo {volume} {116}},\ \bibinfo {pages} {016802} (\bibinfo {year}
  {2016})}\BibitemShut {NoStop}%
\bibitem [{\citenamefont {Lindner}\ \emph
  {et~al.}(2011{\natexlab{a}})\citenamefont {Lindner}, \citenamefont {Refael},\
  and\ \citenamefont {Galitski}}]{Floquetref6}%
  \BibitemOpen
  \bibfield  {author} {\bibinfo {author} {\bibfnamefont {N.~H.}\ \bibnamefont
  {Lindner}}, \bibinfo {author} {\bibfnamefont {G.}~\bibnamefont {Refael}},\
  and\ \bibinfo {author} {\bibfnamefont {V.}~\bibnamefont {Galitski}},\
  }\bibfield  {title} {\bibinfo {title} {Floquet topological insulator in
  semiconductor quantum wells},\ }\href {https://doi.org/10.1038/nphys1926}
  {\bibfield  {journal} {\bibinfo  {journal} {Nature Physics}\ }\textbf
  {\bibinfo {volume} {7}},\ \bibinfo {pages} {490} (\bibinfo {year}
  {2011}{\natexlab{a}})}\BibitemShut {NoStop}%
\bibitem [{\citenamefont {Rechtsman}\ \emph
  {et~al.}(2013{\natexlab{b}})\citenamefont {Rechtsman}, \citenamefont
  {Zeuner}, \citenamefont {Plotnik}, \citenamefont {Lumer}, \citenamefont
  {Podolsky}, \citenamefont {Dreisow}, \citenamefont {Nolte}, \citenamefont
  {Segev},\ and\ \citenamefont {Szameit}}]{Floquetref7}%
  \BibitemOpen
  \bibfield  {author} {\bibinfo {author} {\bibfnamefont {M.~C.}\ \bibnamefont
  {Rechtsman}}, \bibinfo {author} {\bibfnamefont {J.~M.}\ \bibnamefont
  {Zeuner}}, \bibinfo {author} {\bibfnamefont {Y.}~\bibnamefont {Plotnik}},
  \bibinfo {author} {\bibfnamefont {Y.}~\bibnamefont {Lumer}}, \bibinfo
  {author} {\bibfnamefont {D.}~\bibnamefont {Podolsky}}, \bibinfo {author}
  {\bibfnamefont {F.}~\bibnamefont {Dreisow}}, \bibinfo {author} {\bibfnamefont
  {S.}~\bibnamefont {Nolte}}, \bibinfo {author} {\bibfnamefont
  {M.}~\bibnamefont {Segev}},\ and\ \bibinfo {author} {\bibfnamefont
  {A.}~\bibnamefont {Szameit}},\ }\bibfield  {title} {\bibinfo {title}
  {Photonic floquet topological insulators},\ }\href
  {https://doi.org/10.1038/nature12066} {\bibfield  {journal} {\bibinfo
  {journal} {Nature}\ }\textbf {\bibinfo {volume} {496}},\ \bibinfo {pages}
  {196} (\bibinfo {year} {2013}{\natexlab{b}})}\BibitemShut {NoStop}%
\bibitem [{\citenamefont {Perez-Piskunow}\ \emph
  {et~al.}(2014{\natexlab{c}})\citenamefont {Perez-Piskunow}, \citenamefont
  {Usaj}, \citenamefont {Balseiro},\ and\ \citenamefont
  {Torres}}]{Floquetref9}%
  \BibitemOpen
  \bibfield  {author} {\bibinfo {author} {\bibfnamefont {P.~M.}\ \bibnamefont
  {Perez-Piskunow}}, \bibinfo {author} {\bibfnamefont {G.}~\bibnamefont
  {Usaj}}, \bibinfo {author} {\bibfnamefont {C.~A.}\ \bibnamefont {Balseiro}},\
  and\ \bibinfo {author} {\bibfnamefont {L.~E. F.~F.}\ \bibnamefont {Torres}},\
  }\bibfield  {title} {\bibinfo {title} {Floquet chiral edge states in
  graphene},\ }\href {https://doi.org/10.1103/PhysRevB.89.121401} {\bibfield
  {journal} {\bibinfo  {journal} {Phys. Rev. B}\ }\textbf {\bibinfo {volume}
  {89}},\ \bibinfo {pages} {121401} (\bibinfo {year}
  {2014}{\natexlab{c}})}\BibitemShut {NoStop}%
\bibitem [{\citenamefont {Afzal}\ \emph
  {et~al.}(2020{\natexlab{b}})\citenamefont {Afzal}, \citenamefont
  {Zimmerling}, \citenamefont {Ren}, \citenamefont {Perron},\ and\
  \citenamefont {Van}}]{Floquetref11}%
  \BibitemOpen
  \bibfield  {author} {\bibinfo {author} {\bibfnamefont {S.}~\bibnamefont
  {Afzal}}, \bibinfo {author} {\bibfnamefont {T.~J.}\ \bibnamefont
  {Zimmerling}}, \bibinfo {author} {\bibfnamefont {Y.}~\bibnamefont {Ren}},
  \bibinfo {author} {\bibfnamefont {D.}~\bibnamefont {Perron}},\ and\ \bibinfo
  {author} {\bibfnamefont {V.}~\bibnamefont {Van}},\ }\bibfield  {title}
  {\bibinfo {title} {Realization of anomalous floquet insulators in strongly
  coupled nanophotonic lattices},\ }\href
  {https://doi.org/10.1103/PhysRevLett.124.253601} {\bibfield  {journal}
  {\bibinfo  {journal} {Phys. Rev. Lett.}\ }\textbf {\bibinfo {volume} {124}},\
  \bibinfo {pages} {253601} (\bibinfo {year} {2020}{\natexlab{b}})}\BibitemShut
  {NoStop}%
\bibitem [{\citenamefont {Lindner}\ \emph
  {et~al.}(2011{\natexlab{b}})\citenamefont {Lindner}, \citenamefont {Refael},\
  and\ \citenamefont {Galitski}}]{Floref2}%
  \BibitemOpen
  \bibfield  {author} {\bibinfo {author} {\bibfnamefont {N.~H.}\ \bibnamefont
  {Lindner}}, \bibinfo {author} {\bibfnamefont {G.}~\bibnamefont {Refael}},\
  and\ \bibinfo {author} {\bibfnamefont {V.}~\bibnamefont {Galitski}},\
  }\bibfield  {title} {\bibinfo {title} {Floquet topological insulator in
  semiconductor quantum wells},\ }\href {https://doi.org/10.1038/nphys1926}
  {\bibfield  {journal} {\bibinfo  {journal} {Nature Physics}\ }\textbf
  {\bibinfo {volume} {7}},\ \bibinfo {pages} {490} (\bibinfo {year}
  {2011}{\natexlab{b}})}\BibitemShut {NoStop}%
\bibitem [{\citenamefont {Cayssol}\ \emph {et~al.}()\citenamefont {Cayssol},
  \citenamefont {Dóra}, \citenamefont {Simon},\ and\ \citenamefont
  {Moessner}}]{Floref3}%
  \BibitemOpen
  \bibfield  {author} {\bibinfo {author} {\bibfnamefont {J.}~\bibnamefont
  {Cayssol}}, \bibinfo {author} {\bibfnamefont {B.}~\bibnamefont {Dóra}},
  \bibinfo {author} {\bibfnamefont {F.}~\bibnamefont {Simon}},\ and\ \bibinfo
  {author} {\bibfnamefont {R.}~\bibnamefont {Moessner}},\ }\bibfield  {title}
  {\bibinfo {title} {Floquet topological insulators},\ }\href
  {https://doi.org/https://doi.org/10.1002/pssr.201206451} {\bibfield
  {journal} {\bibinfo  {journal} {physica status solidi (RRL) – Rapid
  Research Letters}\ }\textbf {\bibinfo {volume} {7}},\ \bibinfo {pages}
  {101}}\BibitemShut {NoStop}%
\bibitem [{\citenamefont {Liu}\ \emph {et~al.}(2019{\natexlab{b}})\citenamefont
  {Liu}, \citenamefont {Shabani},\ and\ \citenamefont {Mitra}}]{Majref6}%
  \BibitemOpen
  \bibfield  {author} {\bibinfo {author} {\bibfnamefont {D.~T.}\ \bibnamefont
  {Liu}}, \bibinfo {author} {\bibfnamefont {J.}~\bibnamefont {Shabani}},\ and\
  \bibinfo {author} {\bibfnamefont {A.}~\bibnamefont {Mitra}},\ }\bibfield
  {title} {\bibinfo {title} {Floquet majorana zero and $\ensuremath{\pi}$ modes
  in planar josephson junctions},\ }\href
  {https://doi.org/10.1103/PhysRevB.99.094303} {\bibfield  {journal} {\bibinfo
  {journal} {Phys. Rev. B}\ }\textbf {\bibinfo {volume} {99}},\ \bibinfo
  {pages} {094303} (\bibinfo {year} {2019}{\natexlab{b}})}\BibitemShut
  {NoStop}%
\bibitem [{\citenamefont {Peng}\ \emph {et~al.}(2021)\citenamefont {Peng},
  \citenamefont {Haim}, \citenamefont {Karzig}, \citenamefont {Peng},\ and\
  \citenamefont {Refael}}]{Majref6a}%
  \BibitemOpen
  \bibfield  {author} {\bibinfo {author} {\bibfnamefont {C.}~\bibnamefont
  {Peng}}, \bibinfo {author} {\bibfnamefont {A.}~\bibnamefont {Haim}}, \bibinfo
  {author} {\bibfnamefont {T.}~\bibnamefont {Karzig}}, \bibinfo {author}
  {\bibfnamefont {Y.}~\bibnamefont {Peng}},\ and\ \bibinfo {author}
  {\bibfnamefont {G.}~\bibnamefont {Refael}},\ }\bibfield  {title} {\bibinfo
  {title} {Floquet majorana bound states in voltage-biased planar josephson
  junctions},\ }\bibfield  {journal} {\bibinfo  {journal} {Physical Review
  Research}\ }\textbf {\bibinfo {volume} {3}},\ \href
  {https://doi.org/10.1103/physrevresearch.3.023108}
  {10.1103/physrevresearch.3.023108} (\bibinfo {year} {2021})\BibitemShut
  {NoStop}%
\bibitem [{\citenamefont {Seetharam}\ \emph {et~al.}(2015)\citenamefont
  {Seetharam}, \citenamefont {Bardyn}, \citenamefont {Lindner}, \citenamefont
  {Rudner},\ and\ \citenamefont {Refael}}]{occu_PhysRevX}%
  \BibitemOpen
  \bibfield  {author} {\bibinfo {author} {\bibfnamefont {K.~I.}\ \bibnamefont
  {Seetharam}}, \bibinfo {author} {\bibfnamefont {C.-E.}\ \bibnamefont
  {Bardyn}}, \bibinfo {author} {\bibfnamefont {N.~H.}\ \bibnamefont {Lindner}},
  \bibinfo {author} {\bibfnamefont {M.~S.}\ \bibnamefont {Rudner}},\ and\
  \bibinfo {author} {\bibfnamefont {G.}~\bibnamefont {Refael}},\ }\bibfield
  {title} {\bibinfo {title} {Controlled population of floquet-bloch states via
  coupling to bose and fermi baths},\ }\href
  {https://doi.org/10.1103/PhysRevX.5.041050} {\bibfield  {journal} {\bibinfo
  {journal} {Phys. Rev. X}\ }\textbf {\bibinfo {volume} {5}},\ \bibinfo {pages}
  {041050} (\bibinfo {year} {2015})}\BibitemShut {NoStop}%
\bibitem [{\citenamefont {Li}\ \emph {et~al.}(2014)\citenamefont {Li},
  \citenamefont {Kundu}, \citenamefont {Zhong},\ and\ \citenamefont
  {Seradjeh}}]{MajsigAK}%
  \BibitemOpen
  \bibfield  {author} {\bibinfo {author} {\bibfnamefont {Y.}~\bibnamefont
  {Li}}, \bibinfo {author} {\bibfnamefont {A.}~\bibnamefont {Kundu}}, \bibinfo
  {author} {\bibfnamefont {F.}~\bibnamefont {Zhong}},\ and\ \bibinfo {author}
  {\bibfnamefont {B.}~\bibnamefont {Seradjeh}},\ }\bibfield  {title} {\bibinfo
  {title} {Tunable floquet majorana fermions in driven coupled quantum dots},\
  }\href {https://doi.org/10.1103/PhysRevB.90.121401} {\bibfield  {journal}
  {\bibinfo  {journal} {Phys. Rev. B}\ }\textbf {\bibinfo {volume} {90}},\
  \bibinfo {pages} {121401} (\bibinfo {year} {2014})}\BibitemShut {NoStop}%
\bibitem [{\citenamefont {Matsyshyn}\ \emph {et~al.}(2023)\citenamefont
  {Matsyshyn}, \citenamefont {Song}, \citenamefont {Villadiego},\ and\
  \citenamefont {Shi}}]{matsyshyn2023fermi}%
  \BibitemOpen
  \bibfield  {author} {\bibinfo {author} {\bibfnamefont {O.}~\bibnamefont
  {Matsyshyn}}, \bibinfo {author} {\bibfnamefont {J.~C.}\ \bibnamefont {Song}},
  \bibinfo {author} {\bibfnamefont {I.~S.}\ \bibnamefont {Villadiego}},\ and\
  \bibinfo {author} {\bibfnamefont {L.-k.}\ \bibnamefont {Shi}},\ }\bibfield
  {title} {\bibinfo {title} {The fermi-dirac staircase occupation of floquet
  bands and current rectification inside the optical gap of metals: a rigorous
  perspective},\ }\href@noop {} {\bibfield  {journal} {\bibinfo  {journal}
  {arXiv preprint arXiv:2301.00811}\ } (\bibinfo {year} {2023})}\BibitemShut
  {NoStop}%
\bibitem [{\citenamefont {Rodriguez-Vega}\ \emph {et~al.}(2018)\citenamefont
  {Rodriguez-Vega}, \citenamefont {Lentz},\ and\ \citenamefont
  {Seradjeh}}]{Floq_pert}%
  \BibitemOpen
  \bibfield  {author} {\bibinfo {author} {\bibfnamefont {M.}~\bibnamefont
  {Rodriguez-Vega}}, \bibinfo {author} {\bibfnamefont {M.}~\bibnamefont
  {Lentz}},\ and\ \bibinfo {author} {\bibfnamefont {B.}~\bibnamefont
  {Seradjeh}},\ }\bibfield  {title} {\bibinfo {title} {Floquet perturbation
  theory: formalism and application to low-frequency limit},\ }\href
  {https://doi.org/10.1088/1367-2630/aade37} {\bibfield  {journal} {\bibinfo
  {journal} {New Journal of Physics}\ }\textbf {\bibinfo {volume} {20}},\
  \bibinfo {pages} {093022} (\bibinfo {year} {2018})}\BibitemShut {NoStop}%
\end{thebibliography}%
	\renewcommand{\theequation}{S\arabic{equation}}
	\setcounter{equation}{0}
	\renewcommand{\thefigure}{S\arabic{figure}}
	\setcounter{figure}{0}
	\appendix{}
\section*{Appendices}
	\subsection*{A. Driven quantum systems}\label{sec:driven}
	This section provides further details of the transport simulation used in the main text, based on Floquet Green's function techniques. Let us consider the Hamiltonian of a generic periodically driven system that is connected to reservoirs, written in the Bogolibov-de-Gennes basis, as
	\begin{align}\label{hamlsyslead}
		& H_{\text{S}} = \frac12\sum_{x,x',\eta,\eta'} \Psi^{\dagger}_{x\eta}(t)\,h^S_{x\eta,x'\eta'}(t)\Psi_{x'\eta'}(t),\\
		& H_{\lambda} = \frac12\sum_{y_{\lambda},y_{\lambda}',\eta,\eta'} \Phi^{\dagger \lambda}_{y_{\lambda}\eta}(t)\,h^{\lambda}_{y\eta,y_{\lambda}'\eta'}(t)\Phi^{\lambda}_{y'\eta'}(t).
	\end{align}
	The system and $\lambda$-th reservoir Hamiltonians are denoted by $H_S$ and $H_{\lambda}$, respectively. Here, $x, x'$ and $y_{\lambda},y_{\lambda}'$ are the site indices of the system and the reservoirs in the same direction, respectively. $\eta,\eta'$ are particle and hole degrees of freedom at a given site. $N$ is the number of sites in the system. $\Psi^{\dagger}_{x}=(a^{\dagger}_{x},a_{x})^{T}$, $\Phi^{\dagger \lambda}_{y}=(c^{\dagger \lambda}_{y_{\lambda}},c^{\lambda}_{y_{\lambda}})^{T}$, where $a_x$ and $c_{y_{\lambda}}^{\lambda}$ are electronic annihilation operators at site $x$ and $y$, respectively, for the system and the $\lambda$-th reservoir. Tunneling Hamiltonians, which connect the reservoirs with the system, are given by:
	\begin{align}\nonumber\label{hamltunneling}
		& H_{\text{S}\lambda}(t) = \sum_{x\eta,y_{\lambda}\eta'}\frac12 \Big(\Psi^{\dagger}_{x\eta}(t)V^{\lambda}_{x\eta,y_{\lambda}\eta'}\Phi^{\lambda}_{y_{\lambda}\eta'}(t)+\Phi^{\dagger   \lambda}_{y_{\lambda}\eta'}(t)\\
		&\hspace{14mm}V^{\lambda*}_{x\eta,y_{\lambda}\eta'}\Psi_{x\eta}(t)\Big).
	\end{align}
	$V^{\lambda}$ denotes the tunneling matrix that connects the system to the $\lambda$-th reservoir. Using Heisenberg's equation for the evolution of operators, for the elements of $\Phi^{\lambda}_{y_{\lambda}}(t)$, one finds the equation of motion to be,
	\begin{align}
		&\dot{\Phi}^{\lambda}(t)=-i\left(h^{\lambda}\Phi^{\lambda}(t)+V^{\lambda\dagger}\Psi(t)\right),
	\end{align}
	where we have written the above equation in its matrix-valued form. The solution can be written as
	\begin{align}
		&\Phi^{\lambda}(t)=i g^{\lambda}(t-t_0)\Phi^{\lambda}(t_0)+\int_{t_0}^{t}dt' g^{\lambda}(t-t')V^{\dagger \lambda}\Psi(t').
	\end{align}
	Here $t_0$ is the switching time when the reservoir-to-system connection is made, which we assume to be in the distant past, i.e. $t_0 \rightarrow -\infty$. Here, Green’s function of the  $\lambda$th lead is given by
	\begin{align}
		&g^{\lambda}(t,t')=-ie^{-ih^{\lambda}(t-t')}\theta(t-t').
	\end{align}
	Similarly, the equation of motion for the system operators can be written as
	\begin{align}
		&\left(i\mathbb{I}\frac{d}{dt}-h^{S}\right)\Psi(t)- i\int_{t_0}^{t}dt'  \Gamma(t-t')\Psi(t')=
		\sum_{\lambda}V^{\lambda}\xi^{\lambda}(t),\label{eq:eqm}
	\end{align}
	where $\xi^{\lambda}(t)=i g^{\lambda}(t-t_0)\Phi^{\lambda}(t_0)$,
	and the coupling matrix
	\begin{align}
		\Gamma(t-t')=-i\sum_{\lambda} V^{\lambda}[g^{\lambda}(t-t')]V^{\dagger\lambda}.
	\end{align}
	The Green's function of the equation, Eq.~(\ref{eq:eqm}) satisfies,
	\begin{align}
		&\left(i\mathbb{I}\frac{d}{dt}-h^{S}\right)G(t,t')- i\int_{0}^{\infty}d\tau  \Gamma(\tau)G(t-\tau,t')=
		\delta(t-t').\label{eq:GF}
	\end{align}
	In the flat-band limit, the Green's function of the leads,  $g^{\lambda}(\omega)$, is given by $\rho^{\lambda} = -\frac{1}{\pi} \text{Im}[g^{\lambda}(\omega)]$, which is independent of $\omega$, and $\text{Re}[g^{\lambda}(\omega)]=0$. This implies $\Gamma(\tau) = \Gamma \delta(\tau)$, and we can write the Green's function for the eq.~(\ref{eq:eqm}),
	\begin{align}\label{eq:Greens_function_sys_k}\nonumber
		&G^{(k)}(\omega) = \int_0^{T}\frac{dt}{T}e^{ik\Omega t}\int_{0}^{\infty}e^{i\omega \tau}G(t,t-\tau).\\
		&\hspace{12mm}=\sum_{\alpha,n}\frac{|u_{\alpha}^{(n+k)}\ra\la u^{(n)+}_{\alpha}|}{\omega - \epsilon_{\alpha} -n\Omega + i \gamma_{\alpha}},
	\end{align}
	where $\Omega = 2\pi/T$.
	The system operators are then solved using
	\begin{align}
		\Psi(t) &= \sum_{k,\lambda} \int \frac{d\omega}{2\pi} e^{-i\omega t}e^{-ik\Omega t}G^{(k)}(\omega)V^{\lambda}\xi^{\lambda}(\omega)\label{eq:sys-sol-gf}.
	\end{align}
	In our numerical simulations, we have left and right leads indexed by $\lambda=L$ and $R$, respectively. In our case, $\eta$ represents the particle-hole basis. The explicit non-vanishing elements of the tunneling matrices, in the particle-hole basis, are then given by
	\begin{align}
		& V^{L}_{1,1_{L}}=t^{L}\begin{pmatrix}
			1	& 0\\
			0	& -1
		\end{pmatrix},\quad\quad & V^{R}_{N,1_{R}}=t^{R}\begin{pmatrix}
			1	& 0\\
			0	& -1
		\end{pmatrix}.
	\end{align}
	\begin{figure*}
		\includegraphics[width=0.99\textwidth]{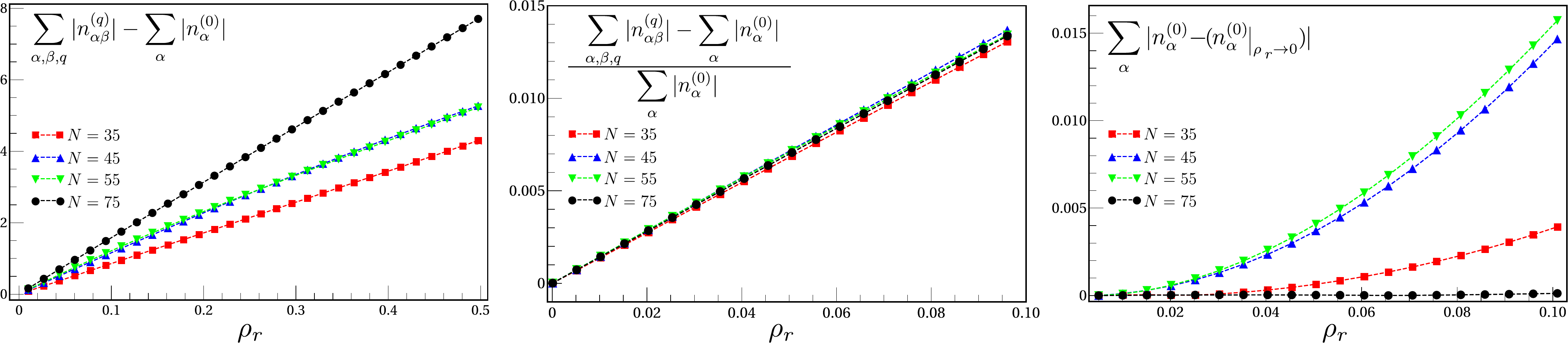}
		\caption{Nature of the occupation probabilities of the Floquet states (Eq.~(\ref{eq:nabq})). For different system sizes, we plot the values of the off-diagonal Fourier components of the occupation number matrix as a function of the tunneling amplitude to the reservoir. Left: The off-diagonal Fourier components of the occupation number matrix go to zero as the tunneling to the leads vanishes. Middle: For various system sizes, the relative values of the off-diagonal Fourier components of the occupation number matrix in comparison to the diagonal, static components of the occupation number matrix. Right: As a function of the tunneling amplitude to the reservoir, the absolute difference between diagonal occupation and diagonal occupations with limit $\rho_L\rightarrow0$. In the limit of small tunneling amplitudes to the reservoir, these results confirm that $n_{\alpha\beta}^{(q)} \approx n_{\alpha}\delta_{\alpha\beta}\delta{q0}$.}
		\label{fig:occuoffdiagonal}
	\end{figure*}
	\noindent Here $t^{\lambda}$ are the tunneling parameters for the connections to reservoirs, which we consider identical. The thermal correlations among the reservoir operators is given by
	\begin{align}
		&\langle \xi^{\lambda\dagger}_{1_{\lambda}\eta}(\omega)\xi^{\lambda'}_{1_{\lambda'}\eta'}(\omega')\rangle = (2\pi)^2\delta_{\lambda\lambda'}\delta_{\eta\eta'}\delta(\omega,\omega') \rho^{\lambda}f^{\lambda}(\omega,\mu^{\lambda},\mathcal{\beta}^{\lambda}),
	\end{align}
	where for $\lambda-$th reservoir, $f^{\lambda}(\omega,\mu^{\lambda},\mathcal{\beta}^{\lambda})$ is the Fermi distribution function, $\mathcal{\beta}=1/\theta^{\lambda}$, $\theta^{\lambda}$ is the temperature, and $\mu^{\lambda}$ is the chemical potential (the reservoirs are identical in our case). The thermal average is taken over the reservoirs' states.	
	
\subsection*{B. Occupations of Floquet states}\label{sec:occupation}
	In this section, we derive a simplified form of the occupation probability of the Floquet states of a periodically driven system in the limit of weak coupling to an external reservoir. In this weak coupling limit, we write the density matrix of the steady-state system in the basis of the Floquet states of the isolated system as
	\begin{align}\label{eq:dm}
		& \hat{\rho}=\frac1T\int_{0}^{T}dt\sum_{\al\bt}n_{\alpha\beta}(t)|u_{\alpha}(t)\ra\la u_{\beta}(t)|.
	\end{align}
	To compute the coefficients $n_{\alpha\beta}(t)$, we write the	creation operator in site basis as, $\Psi^{\dagger}_{\alpha}(t) = \sum_{i,\eta}d^{*}_{\alpha i\eta}(t)\left(\Psi^{\dagger}_{i}(t)\right)_{\eta}$, here $i$- is the site index and $\eta$ is the particle-hole index. $\Psi^{\dagger}_i(t) = (a^{\dagger}_{i}(t),a_{i}(t))$ and $d_{\alpha i\eta}=\langle u_{\alpha}(t)|i,\eta\rangle$. The coefficients $n_{\alpha\beta}(t)$ can then be written as:
	\begin{align}
		& n_{\al\bt}(t)=\Big\la \Psi^{\dagger}_{\al}(t)\Psi_{\bt}(t) \Big\ra_{\text{Lead avg.}}\nonumber
		\\
		&\hspace{6mm}=\sum_{ij,\eta\eta'}d_{\beta j\eta}(t)d^{*}_{\al i\eta'}(t)\la \Psi^{\dagger}_{i\eta'}(t)\Psi_{j\eta}(t)\ra.\nonumber\\
		&\hspace{6mm}=\sum_{\lambda,ij,\eta\eta'}d^*_{\al i\eta'}(t)d_{\bt j\eta}(t)\int_{-\infty}^{\infty} d\omega f^{\lambda}(\omega,\mu_r,\mathcal{\beta})\nonumber
		\\
		&\hspace{10mm} \sum_{mn}\sum_{\xi\xi'}G_{j\eta,m\bar{\eta}}(t,\omega)\mathbb{V}^{\lambda}_{m\bar{\eta},n\xi'}G^{*}_{i\eta',n\xi'}(t,\omega).
	\end{align}
	Performing further simplifications, we obtain
	\begin{align}
		&\sum_{j\eta}d_{\beta j\eta}(t)G_{j\eta,m\bar{\eta}}(t,\omega_1)\nonumber\\
		&=\sum_{p} \Big(e^{ip\Omega(t)}\langle       u_{\beta}^{+\,\,(p)}|m,\bar{\eta}\rangle\Big)\frac{1}{(-p\Omega+\omega-\epsilon_{\beta}+i\gamma_{\beta})},
	\end{align}
	and similarly,
	\begin{align}\nonumber
		&\sum_{i\eta'}d^*_{\al  i\eta'}(t)G^{*}_{i\eta',n\xi'}(t,\omega)\\
		&=\sum_{p} \Big(e^{-ip\Omega(t)}\la n,\xi'|u_{\al }^{+\,\,(p)}\ra\Big)\frac{1}{(-p\Omega+\omega-\epsilon_{\al }-i\gamma_{\al })}.
	\end{align}
	Using the above two equations, one obtains Fourier modes of the coefficients, 
	\begin{align}
		&n^{(q)}_{\al \bt }=\sum_{\lambda k}\int_{-\infty}^{\infty} \frac{\la u_{\bt }^{+\,\,(k)}|\mathbb{V}^{\lambda}|u_{\al }^{+\,\,(k+q)}\ra f^{\lambda}(\omega,\mu^{\lambda},\beta^{\lambda})d\omega}{(\omega-\epsilon^{(k+q)}_{\al }-i\gamma_{\al })(\omega-\epsilon^{(k)}_{\bt }+i\gamma_{\bt })}.\label{eq:occu2}
	\end{align}
	here $\epsilon^{(k)}_{\al}=\epsilon_{\al}+k\Omega$ and $f^{\lambda}(\omega,\mu^{\lambda},\beta^{\lambda})$ is the Fermi distribution of the $\lambda$ th reservoir. Interestingly, we find that in the large system size and weak coupling to the bath limit, 
	\begin{align}\label{eq:occu_q0}
		&n_{\alpha\beta}^{(q)}\approx n_{\alpha}\delta_{\alpha\beta}\delta_{q0}
	\end{align}
	which we demonstrate in  Fig.~\ref{fig:occuoffdiagonal}. This can also be shown in the following way. Expanding the Fermi distribution function as:
	\begin{figure*}[ht]
		\centering
		\includegraphics[width=.95\textwidth]{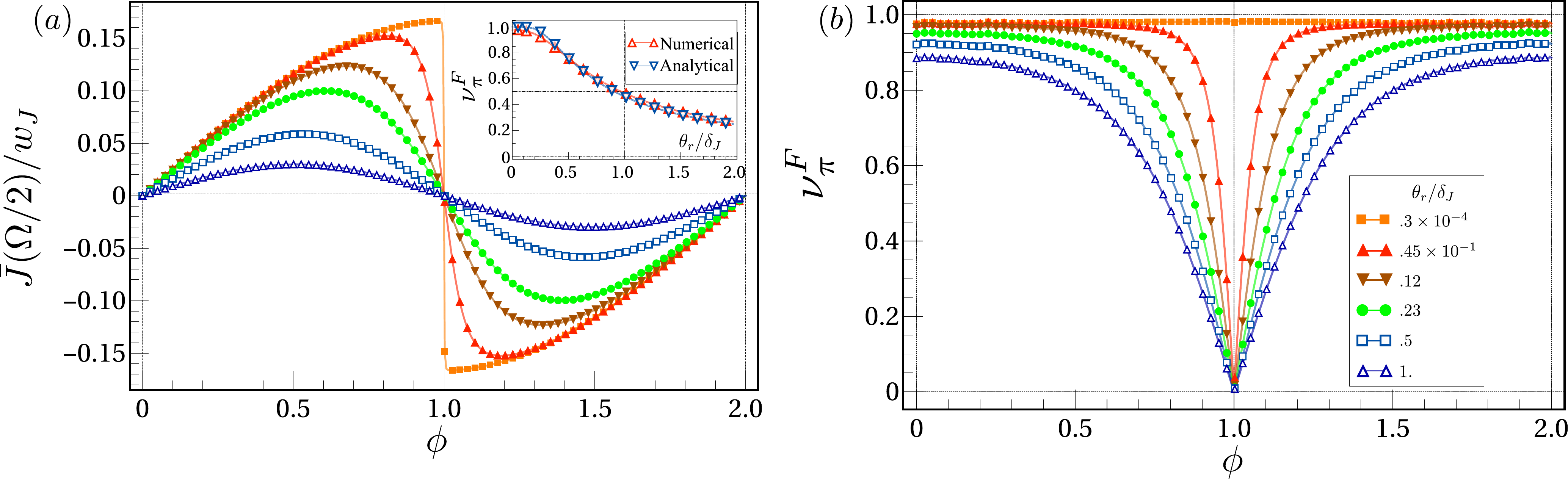}
		\caption{\color{black}{(a) Variation in Josephson current vs phase difference, for different values of reservoir temperature, Here system size is $N=95$ and chemical potential is summed. Inset : Comparison between numerically computed total occupation difference (sum-rule current is computed using Eq.~(\ref{eq:occunum_a})) with the analytical expression given by $\nu^{F}_{\pi}=\tanh\big(\delta_J/2\theta_r\big)-2\delta_J/\Omega$ reservoir temperature (in units of $\delta_{J}(\phi=0)$) for the phase difference $\phi\approxeq .0.498\pi$. (b) Variation in the difference of summed occupation number $\nu^F(\epsilon)$ vs phase difference, here quasi-energies $\epsilon$ are the quasienergies of $\pi$-modes $\Omega/2-\delta_{J}, -\Omega/2+\delta_{J}$ for different reservoir temperatures. Here $\delta_{J}(\phi=0)\approxeq0.0034,\,w_{J}=.01$}}
		\label{fig:fig_temp}
	\end{figure*}
	\begin{align}\nonumber
		& f(\omega,\mu^{\lambda},\mathcal{\beta}^{\lambda})=\frac{1}{e^{\mathcal{\beta}^{\lambda}(\omega-\mu^{\lambda})}}
		\\
		&\hspace{16mm}=-\frac{1}{\mathcal{\beta}^{\lambda}}\sum_{n\in I}\frac{1}{(\omega-\mu^{\lambda})-\frac{(2n+1)i\pi}{\mathcal{\beta}^{\lambda}}}+\frac12,
	\end{align}
	and performing the integrals, we obtain,
	\begin{align}\nonumber\label{eq:nabq}
		n_{\al\bt}^{(q)}&=\sum_{\lambda,k}\frac{ \la u_{\bt }^{+\,\,(k)}|\mathbb{V}^{\lambda}|u_{\al }^{+\,\,(k+q)}\ra}{\epsilon^{(k+q)}_{\al}-\epsilon^{(k)}_{\beta}-i(\gamma_{\al }+\gamma_{\bt})}\Bigg[-i\pi+\nonumber\\
		& \mathbf{\Gamma}_D\Bigg(\frac12+\frac{\mathcal{\beta}^{\lambda}}{2\pi}\xi^{\lambda(k+q)}_{\al}\Bigg)-\mathbf{\Gamma}_D\Bigg(\frac12-\frac{\mathcal{\beta}_r^{\lambda}}{2\pi}\xi^{\lambda(k)}_{\beta}\Bigg)\Bigg]    ,
	\end{align} 	
	where $\xi^{\lambda(k)}_{\al/\beta}=\left(i\epsilon^{(k)}_{\al/\beta}+\gamma_{\al/\beta}-i\mu^{\lambda}\right)$ and $\mathbf{\Gamma}_D(\cdot)$ is the Digamma function. 
	
\begin{figure}[t]
	\centering
	\includegraphics[width=0.45\textwidth]{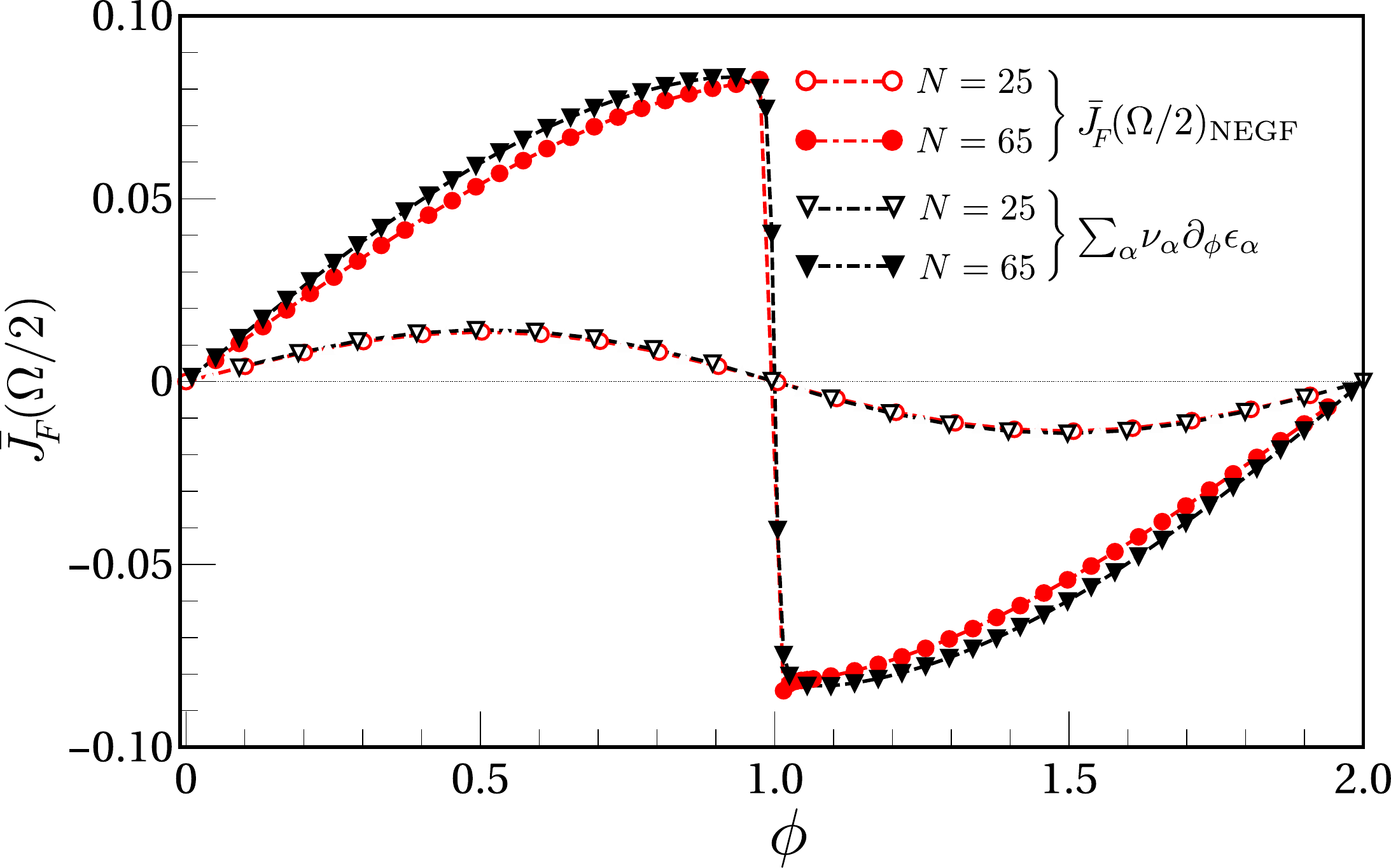}
	\caption{\color{black}{Josephson current calculated from the correlation function (red) and the occupation of quasi-energy modes (black), where the reservoir is connected to every site of each superconductor (in this case, occupation is given by the $\theta-$ function), vs. phase difference for system sizes 25 and 65.}}
	\label{fig:J_corr_J_occu_J_theta_ns_25_ns_65_vs_phir}
\end{figure}

	For the sake of simplicity, suppose identical reservoirs with chemical potential $\mu_r$ and temperature $\theta_r$ are connected at every site of the superconductors with electronic tunneling amplitude $w_r$, then in the limit $w_r \rightarrow0$, $\gamma_{\alpha}\approx \pi \rho_rw_r^{2}\sum_k\la u_{\al }^{(k)}|u_{\al}^{(k)}\ra$, $\rho_r$ being the density of states of the reservoir. In this limit, $n_{\alpha\beta}^{(q)}\approx n_{\alpha}\delta_{\alpha\beta}\delta_{q0} + \mathcal{O}(\gamma)$, which can be written as:
	\begin{align}
		n_{\al}(\mu_r)&=\sum_{k}\Bigg\{\mathbf{\Gamma}_D\Bigg[\frac12+\frac{\mathcal{\beta}_r\xi^{\lambda(k)}_{\al}}{2\pi}\Bigg]-\mathbf{\Gamma}_D\Bigg[\frac12-\frac{\mathcal{\beta}_r\xi^{\lambda(k)}_{\alpha}}{2\pi}\Bigg]\nonumber\\
		&\hspace{12mm}-i\pi\Bigg\}\frac{i\la u_{\alpha }^{(k)}|u_{\al }^{(k)}\ra}{2\pi}\label{eq:occunum_a}\\
		&=\sum_{k} f_{r}(\epsilon_{\al }+k\Omega-\mu_r)\la u_{\al }^{(k)}|u_{\al}^{(k)}\ra.\label{eq:occunum}
	\end{align}
	This allows us to calculate the average of the supercurrent operator within this density matrix. If $H_{\mathcal{S}}(t)$ is the Hamiltonian that contains both the driven superconductors (with a phase  difference of $\phi$), linked via tunneling (except the connections to the reservoir), then the supercurrent operator is defined as $\hat{J}_{\mathcal{S}} = \partial_{\phi}H_{\mathcal{S}}$, and its average in the density matrix is then
	\begin{align}
		& \la\hat{J}_{\mathcal{S}}\ra(\mu_r,t)=\sum_{\al}n_{\al}(\mu_r)\la u_{\al}(t)|\partial_{\phi}H_{\mathcal{S}}|u_{\al}(t)\ra.
	\end{align} 
	Using the relation $H_{\mathcal{S}}|u_{\al}(t)\ra=(\epsilon_{\alpha}+i\partial_t)|u_{\al}(t)\ra$, where $\epsilon_{\alpha}$ are the quasienergies of the full system, we observe
	\begin{align}\nonumber
		& \frac{1}{T}\int_{0}^{T} dt \la u_{\alpha}(t)|(\partial_{\phi}H_{\mathcal{S}})|u_{\alpha}(t)\ra\\\nonumber
		&\hspace{6mm}=\frac{1}{T}\int_{0}^{T}\la u_{\alpha}(t)|\partial_{\phi}\left(H_{\mathcal{S}}|u_{\alpha}(t)\ra\right)-\la u_{\alpha}(t)|H_{\mathcal{S}}|\partial_{\phi}u_{\alpha}(t)\ra\\
		&\hspace{6mm}=\partial_{\phi}\epsilon_{\al}.
	\end{align}
	The time-averaged expectation value of current operator can then be written as:
	\begin{align}
		&\bar{J}(\mu_r)=\la\la\hat{J}_{\mathcal{S}}\ra\ra= \frac{1}{T}\int_{0}^{T}dt\la \hat{J}_{\mathcal{S}}\ra (\mu_r,t)  = \sum_{\al}n_{\al}(\mu_r)\partial_{\phi}\epsilon_{\al},\label{eq:javg}
	\end{align} 
	where the occupation probabilities are given by the Eq.~(\ref{eq:occunum}). Fig.~\ref{fig:J_corr_J_occu_J_theta_ns_25_ns_65_vs_phir} shows a comparison of using this simplified current expression with the tight-binding computation of current Sec.~\ref{sec:Jcorr}, demonstrating a reasonable match between the two methods.
	\subsubsection*{Quantize difference in the occupation of FMF}
	Further simplification can be made for large system sizes when the contribution to the supercurrent is predominantly from the Floquet Majorana bound state (FMF) modes (see Fig:~\ref{fig:E_occu}). The quasi-energy of FMFs localized at the junction is given by $\epsilon_{\alpha}=(\epsilon_b+e^{ib}\delta_{J}),\epsilon_{\bar{\alpha}}=-\epsilon_{\alpha},$ for zero FMF $\epsilon_b=0$, for $\pi$-FMF $\epsilon_b=\Omega/2$, In either of these cases one writes the time averages current (Eq.~(\ref{eq:javg})) as
	\begin{align}\nonumber		       &\bar{J}(\mu_r)=\left(n_{\epsilon_{\alpha}}(\mu_r)\frac{\partial\epsilon_{\alpha}}{\partial{\phi}}+n_{\epsilon_{\bar{\alpha}}}(\mu_r)\frac{\partial\epsilon_{\bar{\alpha}}}{\partial{\phi}}\right)\\\nonumber
		&\hspace{2.5mm}=\left(n_{\epsilon_{\alpha}}(\mu_r)-n_{\epsilon_{\bar{\alpha}}}(\mu_r)\right)\frac{\partial\epsilon_{\alpha}}{\partial{\phi}}\\
		&\hspace{2.5mm}\equiv\nu_{\alpha}(\mu_r)\frac{\partial\epsilon_{\alpha}}{\partial{\phi}}.\label{eq:javg2}
	\end{align}	
	Here $n_{\epsilon_{\alpha}}(\mu_r), n_{\epsilon_{\bar{\alpha}}}(\mu_r)$ are the occupations for $\epsilon_{\alpha},\epsilon_{\bar{\alpha}}$ states respectively, at chemical potential $\mu_r$. The summed difference of occupation probability differences between quasienergy levels $\epsilon_{\alpha}$ and $\epsilon_{\bar{\alpha}}$, at finite temperature, is given by
	\begin{align}	\nonumber                             
		&\nu^{F}_{\alpha}=\lim_{N\rightarrow\infty}\sum_{k=-\infty}^{N}\left(n_{\epsilon_{\alpha}}(\mu_r+k\Omega)-n_{\epsilon_{\bar{\alpha}}}(\mu_r+k\Omega)\right)\\\nonumber
		&=\sum_{m=-\infty}^{\infty}\la u_{\al }^{(m)}|u_{\al}^{(m)}\ra\lim_{N\rightarrow \infty}\sum_{k=-\infty}^{N} \Big[f_{r}(\epsilon_{\al }+m\Omega-\mu_r-k\Omega)\\\nonumber
		&\hspace{35mm}-f_{r}(-\epsilon_{\al }-m\Omega-\mu_r-k\Omega)\Big]\\\nonumber
		&=\sum_{m=-\infty}^{\infty}\la u_{\al }^{(m)}|u_{\al}^{(m)}\ra\Bigg\{\sum_{k=-\infty}^{N-m} \Big[f_{r}(\epsilon_{\al }-\mu_r-k\Omega)\\\nonumber
		&\hspace{45mm}-f_{r}(-\epsilon_{\al }-\mu_r-k\Omega)\Big]\Bigg|_{N\rightarrow\infty}\\
		&\hspace{25mm}-\sum_{k=N-m+1}^{N+m}f_{r}(-\epsilon_{\al }-\mu_r-k\Omega)\Bigg\}\nonumber
	\end{align} 
	For sufficiently large $N$, each summand of the second term is one. Thus, we have
	\begin{align}\nonumber
		&\nu^{F}_{\alpha}=-2\sum_{m=-\infty}^{\infty}m\la u_{\al }^{(m)}|u_{\al}^{(m)}\ra\\
		&+\lim_{N\rightarrow \infty}\sum_{k=-\infty}^{N-m}\Big[f_{r}(\epsilon_{\al }-\mu_r-k\Omega)-f_{r}(-\epsilon_{\al }-\mu_r-k\Omega)\Big]
	\end{align}	
	The first term:
	\begin{align}
		&D_{\al}=\sum_{m=-\infty}^{\infty}m\la u_{\al }^{(m)}|u_{\al}^{(m)}\ra=\frac{i}{2\pi}\int_0^T dt \langle u_{\alpha} (t)|\partial_t|u_{\alpha}(t)\rangle.
	\end{align}
	Using the particle-hole anti-symmetry, where an operator follows, $\{\Gamma,H(t) - i\partial_t\}=0$ (the operator contains a complex conjugation, i.e, $\Gamma i\Gamma^{-1} = -i$), one writes, $|u_{\bar{\alpha}}(t)\rangle = \Gamma |u_{\alpha}(t)\rangle$, implying,  $|u_{\bar{\alpha}}^{(k)}\rangle = \Gamma |u_{\alpha}^{(-k)}\rangle$ (where $\epsilon_{\bar{\alpha}}= - \epsilon_{\alpha}$), and 
	\begin{align}
		D_{\bar{\alpha}} = -D_{\alpha}.
	\end{align}
	Now, using Eq.~(\ref{eq:FlEq}), we write,
	\begin{align}\label{eq:1}
		& D_{\alpha}  = \frac{1}{2\pi}\int_0^T dt \langle u_{\alpha} (t)|(H(t)-\epsilon_{\alpha})|u_{\alpha}(t)\rangle\nonumber\\
		&\hspace{5mm} = \frac{1}{2\pi}\int_0^T dt \langle u_{\alpha} (t)|H(t)|u_{\alpha}(t)\rangle - \frac{\epsilon_{\alpha}}{\Omega}.
	\end{align}
	So, the summed occupation difference now reads as
	\begin{align}\nonumber
		&\nu^{F}_{\alpha} = \frac{2\epsilon_{\alpha}}{\Omega}- \frac{1}{\pi}\int_0^T dt \langle u_{\alpha} (t)|H(t)|u_{\alpha}(t)\rangle+\lim_{N\rightarrow \infty}\\
		&\hspace{4mm}\sum_{k=-\infty}^{N-m}\Big[f_{r}(\epsilon_{\al }-\mu_r-k\Omega)-f_{r}(-\epsilon_{\al }-\mu_r-k\Omega)\Big].
	\end{align}
	For the FMFs, the middle term vanishes, which we can show in the following section. In the limit of large system size, the FMF states are simply superpositions of two unpaired FMF states; these two FMF states are localized at the two ends of a single superconductor. These particle-hole symmetric  states $|u_1(t)\rangle$ and $|u_2(t)\rangle$, follows $|u_1(t)\rangle= \Gamma|u_{1}(t)\ra$ and $|u_{2}(t)\ra=\Gamma|u_{2}(t)\ra$. We construct two orthogonal states as their linear combinations, as
	\begin{align}\nonumber
		&|u_{\al}(t)\ra=\frac{1}{\sqrt{2}}(|u_{1}(t)\ra+\alpha|u_{2}(t)\ra)\\
		&|u_{\bar{\al}}(t)\ra=\frac{1}{\sqrt{2}}(|u_{1}(t)\ra+\alpha^{*}|u_{2}(t)\ra)
	\end{align}
	with the normalization conditions $\la u_{\al }(t)|u_{\al}(t)\ra=1$, $\la u_{\bar{\al}}(t)|u_{\bar{\al}}(t)\ra=1$ and $\la u_{\al }(t)|u_{\bar{\al}}(t)\ra=0,$ for Floquet modes one gets $|\alpha|^2=-1$, implying $\alpha=\pm i$, with this construction,
	\begin{align}\nonumber
		&\la u_{\al }(t)|H(t)|u_{\al}(t)\ra=\la u_{1}(t)|H(t)|u_{1}(t)\ra\\
		&~~~~~~+\la u_{2}(t)|H(t)|u_{2}(t)\ra
		+2\text{Re}[\alpha\la u_{1}(t)|H(t)|u_{2}(t)\ra]
	\end{align}
	The first two terms are zero as the $\{H(t),\Gamma\}=0$, and the third term vanishes at large system sizes as the Hamiltonian is local and the two states are localized at the two ends of the chain. So we arrive at
	\begin{align}\label{eq:diffoccu}\nonumber
		&\nu^{F}_{\alpha} = \lim_{N\rightarrow \infty}\sum_{k=-\infty}^{N-m}\Big[f_{r}(\epsilon_{\al }-\mu_r-k\Omega)-f_{r}(-\epsilon_{\al }-\mu_r-k\Omega)\Big]\\
		&\hspace{10mm}+\frac{2\epsilon_{\alpha}}{\Omega}
	\end{align}

	Now, at a small temperature  $\theta_r\ll \delta_J$, for the 0-FMFs,  $\epsilon_{\al}=-\delta_J,~\mu_b=0,$ Eq.~(\ref{eq:diffoccu}) can be written as:
	\begin{align}\nonumber
		&\nu^{F}_{0} =-\frac{2\delta_J}{\Omega}+\lim_{N\rightarrow \infty}\sum_{k=-\infty}^{N-m}\Big[f_{r}(-\delta_J-k\Omega)-f_{r}(\delta_J-k\Omega)\Big]\\\nonumber
		&\hspace{10mm}=-\frac{2\delta_J}{\Omega}+\Big[f_{r}(-\delta_J)-f_{r}(\delta_J)\Big]\\
		&\hspace{10mm}=-\frac{2\delta_J}{\Omega}+\tanh\left(\frac{\delta_J}{2\theta_r}\right)	
	\end{align}
Similarly for the $\pi$-FMFs, $\epsilon_{\al}=-\Omega/2+\delta_J,~\mu_b=-\Omega/2,$ Eq.~(\ref{eq:diffoccu}) can be written as:
	\begin{align}\nonumber 
		&\nu^{F}_{\pi} =\frac{-\Omega+2\delta_J}{\Omega}+ \lim_{N\rightarrow \infty}\sum_{k=-\infty}^{N-m}\Big[f_{r}(\epsilon_{\al }-\mu_b-k\Omega)\\\nonumber
		&\hspace{10mm}-f_{r}(-\epsilon_{\al }-\mu_b-(k+1)\Omega)\Big]+1\\\nonumber
		&\hspace{4mm}= \frac{2\delta_J}{\Omega}+\Big[f_{r}(\delta_J)-f_{r}(-\delta_J)\Big]\\
		&\hspace{4mm}= \frac{2\delta_J}{\Omega}-\tanh\left(\frac{\delta_J}{2\theta_r}\right)\label{eq:occu_temp}.
	\end{align}	
	
	\subsection*{C. Correlation and bond-current}\label{sec:Jcorr}
	We define the elements of the correlation matrix, between two sites $x$ and $x'$, as
	\begin{align}\nonumber
		&\chi_{x\eta,x'\eta'}(t)=\Big\la \Psi^{\dagger}_{x\eta}(t)\Psi_{x'\eta'}(t) \Big\ra_{\text{Lead average}},
	\end{align}
	where $\eta$, $\eta'$ are indices for the Nambu (particle-hole) basis. In terms of the Floquet Green's functions from the preceding section, we write the time-averaged correlation as
	\begin{align}
		&\bar{\chi}_{x\eta,x'\eta'}=\frac{1}{T}\int_{0}^{T}dt\,\chi_{x\eta,x'\eta'}(t)
		\\\label{eq:correlation_green_function}
		&\hspace{13mm}=\sum_{\lambda,k}\int d\omega   \left(G^{(k)}(\omega)\mathbb{V}^{\lambda}G^{\dagger(k)}(\omega)\right)_{x'\eta',x\eta}f^{\lambda}(\omega),
	\end{align}
	where $\mathbb{V}^{\lambda}= V^{\lambda\dagger}\rho^{\lambda}V^{\lambda}$.  Using the form of the Floquet Green's functions, Eq.~(\ref{eq:Greens_function_sys_k}), the correlation function can be further simplified, at zero temperature limit, as
	\begin{align}\nonumber
		&\bar{\chi}_{x\eta,x'\eta'}=\sum_{\lambda q\alpha\beta km}\Bigg(\frac{\log Z^{(m+q)}_{\alpha}-\log Z^{(m)}_{\beta}}{Z^{(m+q)}_{\alpha}-Z^{(m)}_{\beta}}\Bigg)\la x',\eta'|u_{\alpha}^{(q+k)}\ra
		\\
		&\hspace{15mm}\la u_{\beta}^{(k)} |\eta,x\ra\la u_{\alpha}^{(q+m)^+}|\mathbb{V}^{\lambda}|u_{\beta}^{(m)^{+}}\ra.\nonumber
	\end{align} 
	Here $Z^{(k)}_{\alpha}=\epsilon^{-}_{\alpha}+k\Omega-\mu_{L}$ with $\epsilon_{\alpha}^{\pm}=(\epsilon_{\alpha}\pm i\gamma_{\alpha})$. 
	
	Now, we proceed to derive an expression of bond-current that we use for numerical computation. If $w$ is the hopping amplitude for electron between site $x$ and $x'$, then the hopping Hamiltonian, in the BdG basis is written as 	$h_{x'\rightarrow x}(t)=\sum_{\eta\eta'}\frac12 w\Psi^{\dagger}(t)_{x\eta}\tau^{z}_{\eta\eta'}\Psi(t)_{x'\eta'}+{\rm h.c.}$, where $\tau^z$ is the Pauli matrix in particle-hole space. 
	
		\begin{figure}[t]\label{fig:EPert}
		\centering
		\includegraphics[width=80mm,scale=0.5]{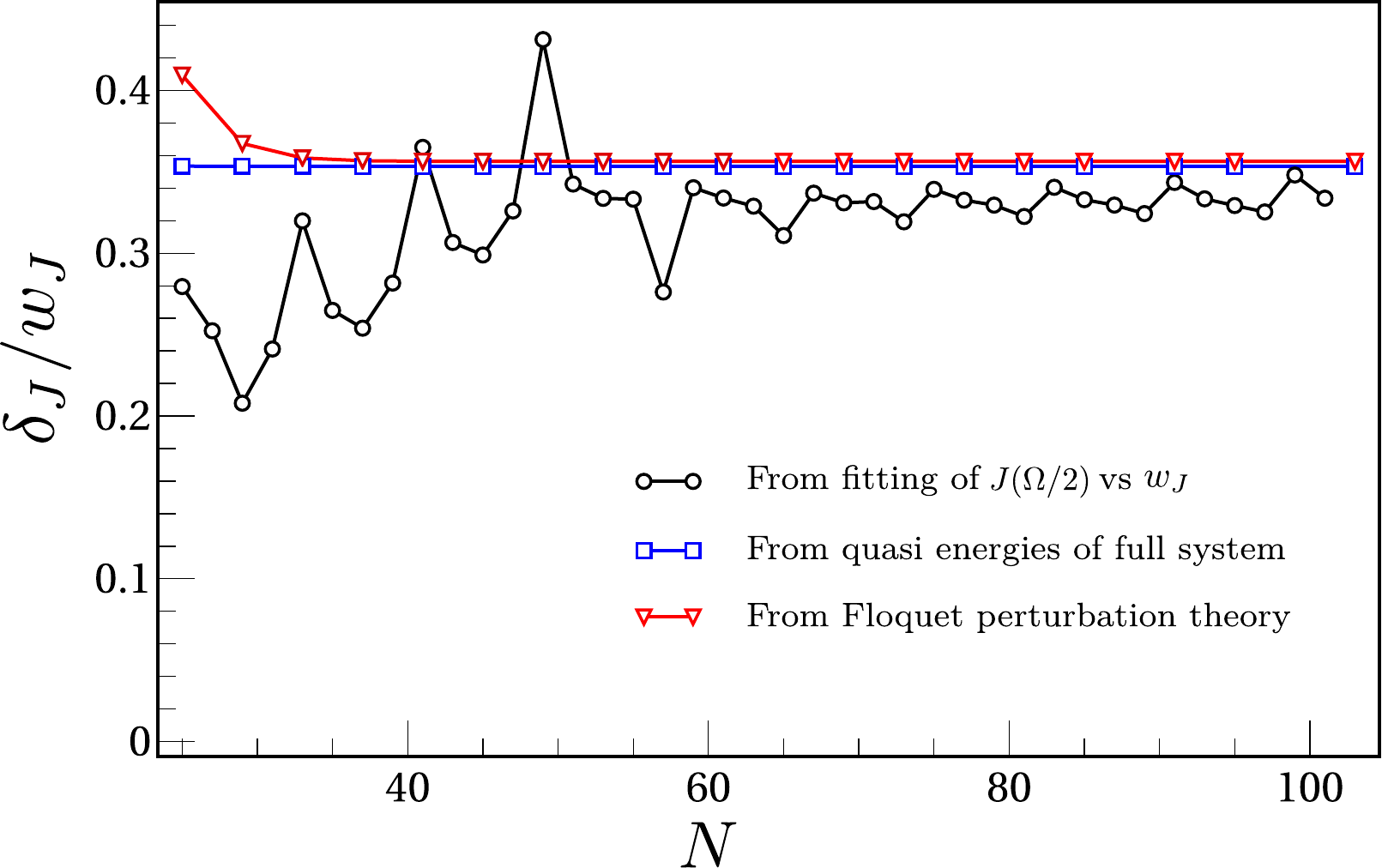}
		\caption{\color{black}{Variation in quasi-energy gaps vs. system size, the gaps are computed (\RNum{1}) from the fitting of Josephson current (using fitting function $J(\Omega/2)=(\tilde{w}_J/2)^2\sin\phi\,E_{J}(\phi)$, where $E_{J}(\phi)=\sqrt{\tilde{w}_J^2\cos^2\phi+E_0^2}$ and $E_0$ is the energy splitting of Majorana bound states in the absence of junction), (\RNum{2}) from the quasi-energies of the full system computed by the evolution operator and (\RNum{3}) from the  Floquet perturbation theory, Appendix D and Ref.~\onlinecite{Floq_pert}.}}
		\label{fig:gapfittingperturbarionplot1rev1}
	\end{figure}
	\begin{figure*}[ht!]
		\includegraphics[width=.99\linewidth]{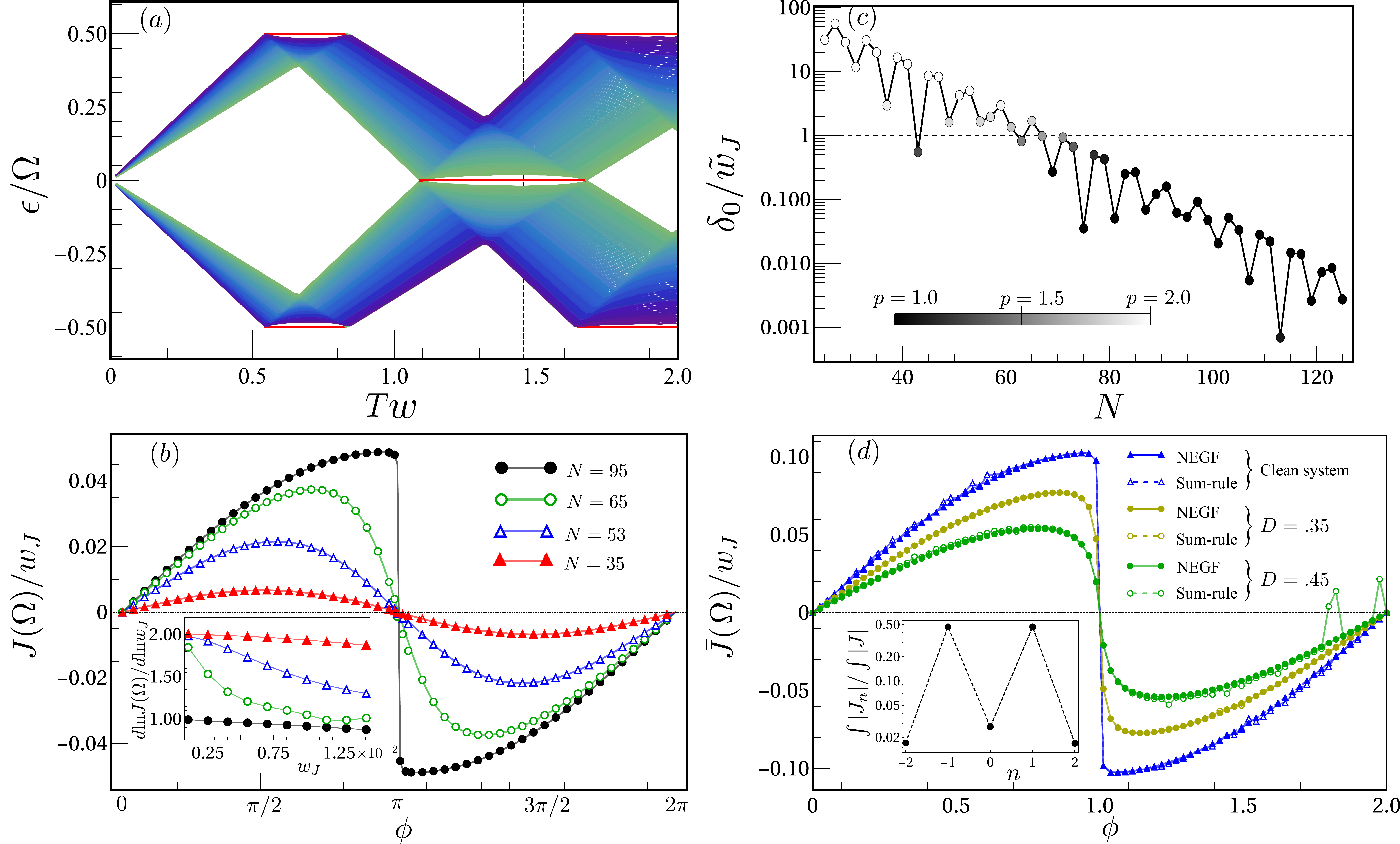}
		\caption{\color{black}{(a) Quasienergy spectrum of the driven Kitaev model of $N=105$ sites, highlighting edge modes (colored red). Here parameters are $\Delta=0.5$, $w=0.5$, $\mu_{0}=4.75$, $\mu_{d}=1.25$. At the time period $T=1.45455$, there are two 0-FMF, which contribute maximally to the Josephson current. (b) Quasienergy gap of 0-FMF localized at end of an uncoupled driven Kitaev chain as a function of the length of the chain ($N$ sites). When $\delta<\tilde{w}_{J}$ ($\tilde{w}_{J}$ is defined in the main text, for the parameter range of the plot, $\tilde{w}_{J} \approx 0.2 w_{J}$), the Josephson current between two such superconductor, $\bar{J}$, is linearly proportional to the $w_{J}$. The grayscale color of the circles at each system size indicates the fitting $\bar{J} \propto w^{p}_{J}$. In the inset we show, in solid line, how $\delta$ varies with system size, which is of the same order of $\delta_0$ (dashed line) for $\delta \gg w_{J}$. $w_{J}=10^{-3}$ and other parameters are the same as in Fig~(a). (c) Current phase relation for various system sizes. For the larger system size the current shows a sharper jump at $\pi$ phase difference same as $\pi-$ Majorana case, a hallmark signature of Majorana fermions. In this limit the Josephson current is linearly proportional to the $w_{J}$ (shown in the inset). Time period of driving $T=1.45454$ and the chemical potential of the external reservoir is set at $\mu_r=\Omega$. (d) Validity of sum-rule for zero 0- case, in presence of static disorder characterized by disorder strength $D$. The summed current is compared with the numerically obtained value using non-equilibrium Green's function method (marked as NEGF), which shows robustness of the sum-rule in this case as well. In the inset we show that the dominant contribution comes from $\mu_r = n\Omega$ with $n=\pm1$, where we show how $\int_0^{2\pi}|\langle \hat{J}(\mu_r = n\Omega) \rangle|d\phi$ as a function of $n$.}}
		\label{fig:J_00}
	\end{figure*}

	The electronic bond current operator can then be defined as $\hat{J}_{x'\rightarrow x}=\frac{\partial h(t,\zeta)_{x'\rightarrow x}}{\partial\zeta}\Big|_{\zeta=0}$, where $h(t,\zeta)=\sum_{\eta\eta'}\frac12 w e^{-i\zeta}\Psi^{\dagger}(t)_{x\eta}\tau^{z}_{\eta\eta'}\Psi(t)_{x'\eta'}+{\rm h.c.}$.  Lead and time averaged bond current from site $x'$ to site $x$, is then defined as 
	\begin{align}\label{eq:Jcorr}
		\bar{J}_{x'\rightarrow x}=-\sum_{\eta\eta'}{\rm Im}\left[w\tau^{z}_{\eta\eta'}\bar{\chi}_{x\eta,x'\eta'}\right].
	\end{align}
	This is the expression we use to calculate the Josephson current numerically, where $x$ and $x'$ are the last and first sites of the left and right superconductors, respectively, which, in our numerical simulation, are linked by a hopping amplitude $w_J$.
	
	\subsubsection*{Numerical results at finite temperature}
	The results for the zero-temperature case are shown in the main text. We plot the Josephson current at finite temperature in Fig.~\ref{fig:fig_temp}a. According to Fig.~\ref{fig:fig_temp}b, the difference in the occupation of 0-FMF changes depending on the temperature. Comparison of the occupation difference determined analytically and numerically in inset of Fig.~\ref{fig:fig_temp}a.
	
%
%

	\subsection*{D. Floquet perturbation in the extended-zone}
	The Shr\"{o}dinger's equation of any periodically driven Hamiltonian $H(t)$ is given as
	\begin{align}
		&i\partial_{t}|\Psi_{\al}(t)\rangle=H(t)|\Psi_{\al}(t)\rangle\label{sch_eom_sys1},
	\end{align}
	where $|\Psi_{\al}(t)\ra=e^{-i\epsilon_{\al}}|u_{\al}(t)\rangle$, with quasienergy $\epsilon_{\al}$ and the time-periodic Floquet-state $|u_{\al}(t)\rangle$. The Shr\"{o}dinger's equation in terms of the Floquet states reads
	\begin{align}\label{eq:FlEq}
		(H(t) - i\partial_t)|u_{\alpha}(t)\rangle = \epsilon_{\alpha}|u_{\alpha}(t)\rangle.
	\end{align}
	In the Fourier-space the above can be written as
	\begin{align}
		\Rightarrow\sum_{p}\left(\delta_{np}n\Omega  -    H^{(n-p)}\right)|u^{(p)}_{\al}\rangle=\epsilon_{\al}|u^{(n)}_{\al}\rangle,
	\end{align}
	The left-hand side matrix is the (static) extended-zone (EZ) Hamiltonian. The eigen-energies of this Hamiltonian are periodic, with the driving frequency's period $\Omega = 2\pi/T$, and the $n$-th floquet zone is defined as $n\Omega - \frac{\Omega}{2}$ to $n\Omega + \frac{\Omega}{2}$. The energy $\epsilon_\alpha + m\Omega$ that lives in the $m$th Floquet zone corresponds to an eigenvector, which is a column vector of the form:
	\begin{align}\label{eq:uEZ}
		|u_{m\alpha}^{\text{EZ}}\rangle =\left(\begin{array}{c}
			\cdot \\
			\cdot \\
			|u_{\alpha}^{(1+m)}\rangle\\
			|u_{\alpha}^{(0+m)}\rangle\\
			|u_{\alpha}^{(-1+m)}\rangle\\
			\cdot \\
			\cdot \\
		\end{array}\right).
	\end{align}
	For a driven topological superconductor, if it hosts 0-FMF, there are nearly-degenerate states (at the two edges of the wire) at energy $n\Omega$ of the above EZ Hamiltonian, whereas, if it hosts $\pi$-FMF, there are degenerate states at the boundary of the Floquet zones (i.e, at energies $(n+1/2)\Omega$ of the EZ Hamiltonian). 

	When there are two such driven topological superconductors, the spectrum of the net EZ Hamiltonian contains \textit{four} edge-states (in the limit of large sizes of the superconductors), two at the far ends of the superconductors and two at the junction of the SC's. In presence of a weak tunnel coupling between the superconductors, one can the perform a degenerate perturbation theory in obtaining a gap between the Majoranas that live at the junction. We plot this perturbative result of the gap in  Fig.~(\ref{fig:gapfittingperturbarionplot1rev1}) and compare with numerical result.
	\vspace{5mm}
	\subsection*{E. 0-FMF case}
	In the Fig.~(\ref{fig:J_00}) we summarize numerical results for the case of a parameters when we have 0-FMF.	
	
\end{document}